\documentclass[12pt,tightenlines,nofootinbib]{revtex4}
\usepackage{amsfonts,amsmath,amssymb}
\usepackage{enumitem}
\usepackage{txfonts}
\usepackage{graphicx,color}
\usepackage{wrapfig,lipsum,booktabs}
\usepackage{floatrow}
\usepackage{hyperref}
\usepackage[normalem]{ulem}
\usepackage{mathrsfs}
\usepackage{mathtools}

\DeclareMathOperator{\sk}{sk}

\DeclarePairedDelimiter\floor{\lfloor}{\rfloor}
\DeclareMathOperator{\R5}{\textsf{RCC5}}

\begin{document}

\title{Spatial representability of neuronal activity}
\author{D. Akhtiamov$^1$, A. G. Cohn$^{2-6}$, Y. Dabaghian$^7$}
\affiliation{$^1$ Einstein institute of Mathematics, The Hebrew University, Jerusalem,  Israel, 9190401 danil.akhtiamov@mail.huji.ac.il \\
$^2$ School of Computing, University of Leeds, Woodhouse Lane, Leeds LS9 2JT, UK a.g.cohn@leeds.ac.uk\\
$^3$ Luzhong Institute of Safety, Environmental Protection Engineering and Materials, Qingdao University of Science \& Technology, Zibo 255000, China\\
$^4$ School of Mechanical and Electrical Engineering, Qingdao University of Science and Technology, Qingdao 260061, China\\
$^5$ Department of Computer Science and Technology, Tongji University, Shanghai 211985, China\\
$^6$ School of Civil Engineering, Shandong University, Jinan 250061, China\\
$^7$ Department of Neurology, The University of Texas McGovern Medical School, 6431 Fannin St, Houston, TX 77030\\
$^{*}$e-mail: Yuri.A.Dabaghian@uth.tmc.edu}
\date{\today}

\begin{abstract}
A common approach to interpreting spiking activity is based on identifying the firing fields---regions
in physical or configuration spaces that elicit responses of neurons. Common examples include hippocampal
place cells that fire at preferred locations in the navigated environment, head direction cells that fire
at preferred orientations of the animal's head, view cells that respond to preferred spots in the visual 
field, etc. In all these cases, firing fields were discovered empirically, by trial and error. We argue 
that the existence and a number of properties of the firing fields can be established theoretically, 
through topological analyses of the neuronal spiking activity.
\end{abstract}

\maketitle
\newpage

\section{Introduction and the physiological background}
\label{sec:intro}

Physiological mechanisms underlying the brain's ability to process spatial information are discovered by
relating parameters of neuronal spiking with characteristics of the external world. In many cases, it is
possible to link neuronal activity to geometric or topological aspects of a certain space---either physical
or auxiliary. For example, a key insight into neuronal computations implemented by the mammalian hippocampus
is due to O'Keefe and Dostrovsky's discovery of a correlation between the firing rate of principal neurons in
rodents' hippocampi and the animals' spatial location \cite{Dostrovsky,OKeefe,Vinogradova}. This discovery 
allowed interpreting these neurons' spiking activity, henceforth called \textit{place cells}, as representations
of spatial domains---their respective \textit{place fields} (Fig.~\ref{fig:fields}A, \cite{Best}). 
\footnote{Throughout the text, terminological definitions are given in \textit{italics}.} 
It then became possible to use place field layout in the navigated environment $\mathcal{E}$---the \textit{place
field map} $M_{\mathcal{E}}$---to decode the animal's ongoing location \cite{Brown1,Barbieri1,Jensen1,Guger},
and even to interpret the place cells' off-line activity during quiescent stages of behavior or in sleep 
\cite{Karlsson1,Wu,Ji,Johnson,Dragoi,Pfeiffer1}, which define our current understanding of the hippocampus'
contribution to spatial awareness \cite{MosMc,Kropff,Derdikman,Grieves}.

In the 90s, a similar line of arguments was applied to cells discovered in rat's postsubiculum and in other 
parts of the brain \cite{Taube,TaubeGood,Wiener}, which fire at a particular orientation of the animal's head. 
The angular domains where such \textit{head direction cells} become active can be viewed as one-dimensional 
($1D$) \textit{head direction fields} in the circular space of planar directions, $S^1$---in direct analogy
with the hippocampal place fields in the navigated space (Fig.~\ref{fig:fields}B). The corresponding \textit{
	head direction map}, $M_{S^1}$, defines the order in which the head direction cells spike during the rat's
movements and the role of these cells in spatial orientation \cite{TaubeGood,Wiener,Savelli}. 
Recently, place cells and head directions cells were discovered in bats' hippocampi; in contrast with rodents
who navigate two-dimensional ($2D$) surfaces (see however \cite{Knierim,Hayman,Jeffery,Griev3D}), bat's 
voluminous place fields cover three-dimensional ($3D$) environments and their head direction fields cover $2D$
tori \cite{Rubin,Finkelstein}. 

The \textit{spatial view cells}, discovered in the late 90s, activate when a primate is looking at their 
preferred spots in the environment (Fig.~\ref{fig:fields}C), regardless of the head direction or location 
\cite{Georges,Rolls1,Rolls2}. Correlating these cells' spike timing with the positions of the \textit{view 
	fields} helped understanding mechanisms of storing and retrieving episodic memories, remembering object
locations, etc. \cite{Buffalo,Araujo}. The principles of information processing in sensory and somatosensory
cortices were also deciphered in terms of receptive fields---domains in sensory spaces, whose stimulation 
elicits in spiking responses of the corresponding neurons \cite{Hubel,Arun,Aertsen,Atencio,Gosselin,DeAngelis}.

In all these cases, referencing an individual neuron's activity to a particular domain in a suitable 
\textit{representing space} $X$ \cite{SchemaS} is key for understanding its contribution and for reasoning
about functions of neuronal ensembles in terms of the corresponding ``maps" \cite{Derdikman,Grieves,Kropff}. 
This raises a natural question: when is a ``spatial" interpretation of neuronal activity at all possible, i.e.,
when there might exist a correspondence between the patterns of neuronal activity and regions in low-dimensional
space?

\section{Approach}
\label{sec:app}

\textbf{A mathematical perspective} on this question is suggested by the simplicial topology framework 
\cite{Alexandrov,Hatcher}. Specifically, if a combination of coactive cells, $c_{i_0},c_{i_1},\ldots,c_{i_k}$
is represented by an abstract \textit{coactivity simplex} (for definitions see Sec.~\ref{sec:met})
\begin{equation}
\sigma_i = [c_{i_0},c_{i_1},\ldots,c_{i_k}],
\label{sigma}
\end{equation}
then the net pool of coactivities observed by the time $t$ forms a simplicial complex 
\begin{equation}
\mathcal{T}(t)=\cup_{i}\sigma_i.
\label{complex}
\end{equation}
On the other hand \cite{Alexandroff,Cech,Hatcher,Edwards}, a similar construction can be carried out for a 
space $X$ covered by a set of regions $\upsilon_i$,
\begin{equation}
X=\cup_i\upsilon_i.
\label{cover}
\end{equation}
If each nonempty overlap between these regions,
\begin{equation}
\upsilon_{\sigma_i}\equiv\upsilon_{i_0}\cap\upsilon_{i_1}\cap\ldots\cap\upsilon_{i_k}\neq\varnothing,
\label{overlap}
\end{equation} 
is formally represented by an abstract simplex, 
\begin{equation}
\nu_{\sigma_i}=[\upsilon_{i_0},\upsilon_{i_1},\ldots,\upsilon_{i_k}],
\label{nsimplex}
\end{equation}
then the cover (\ref{cover}) generates another simplicial complex, known as its \textit{\v{C}ech} or 
\textit{nerve} complex
\begin{equation}
\mathcal{N}_X=\cup_i\nu_{\sigma_i},
\label{nerve}
\end{equation}
which is a spatial analogue of the coactivity complex (\ref{complex}). 
The idea is hence the following: if there is a correspondence between neurons' spiking and spatial regions,
then multi-cell coactivities can be viewed as representations of their firing fields' overlaps \cite{Ghrist1,
Curto,PLoS}. Thus, the question whether a given pool of neuronal activity corresponds to a spatial map can
be answered by verifying \textit{representability} of the corresponding coactivity complex $\mathcal{T}(t)$,
i.e., testing whether the latter has a structure of a nerve $\mathcal{N}_X$ of some cover in a low-dimensional
representing space $X$.

\textbf{Implementation}. As it turns out, representable simplicial complexes exhibit several characteristic
properties that distinguish them among generic simplicial complexes \cite{TancerSur,Tancer2}. Verifying
these properties over biologically relevant $1D$, $2D$ and $3D$ representing spaces is a tractable problem
\cite{TancerSur,Kratochvil}, although exact algorithms for performing such a verification are not known---only
in $1D$ are some methods available \cite{Fulkerson,Kratsch,Habib,Golumbic,Fishburn}. 
Nevertheless, there exist explicit criteria that allow limiting the dimensionality of the representing space
$X$ and eliminating manifestly non-representable complexes based on their homologies, combinatorics of
simplexes and other intrinsic topological properties, which will be used below.

\begin{figure} 
	\includegraphics[scale=0.84]{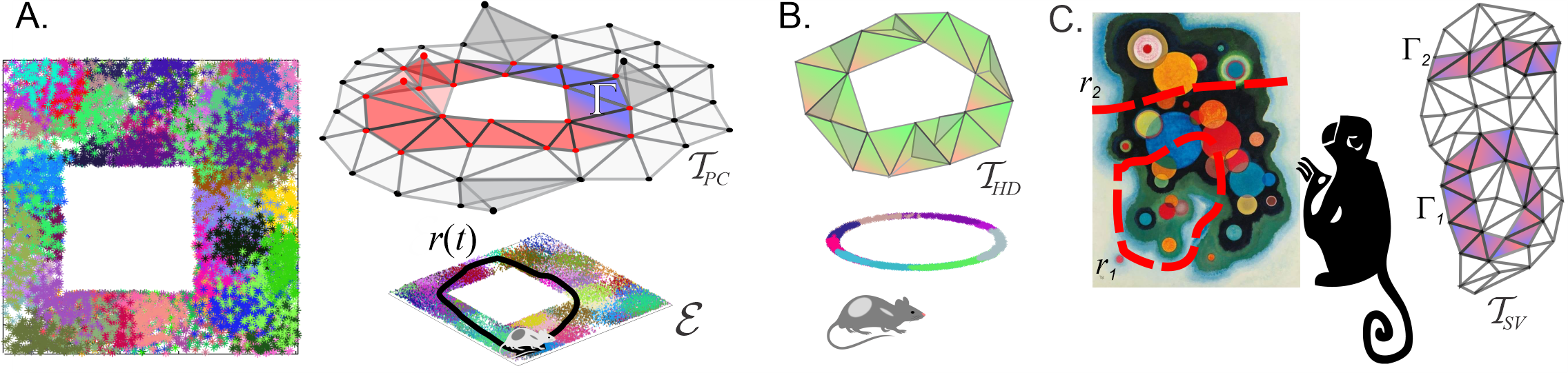} 
	\caption{\footnotesize \textbf{Spatial maps}.
		({\bf A}) A simulated place field map of a small ($1m\times 1m$) environment $\mathcal{E}$, similar
		to the arenas used in typical electrophysiological experiments \cite{Hafting,BrunG}. Dots represent
		spikes produced by the individual cells (color-coded); their locations mark the rat's position at the
		time of spiking. The pool of place cell coactivities is schematically represented by a coactivity 
		complex $\mathcal{T}_{PC}$ (top right). The navigated trajectory $r(t)$ induces a sequence of activated
		simplexes---a simplicial path $\Gamma\in\mathcal{T}_{PC}$.
		({\bf B}) The head direction cell combinations ignited during navigation induce a coactivity complex
		$\mathcal{T}_{HD}$ (top). The corresponding head direction fields cover a unit circle---the space of
		directions (bottom). 
		({\bf C}) Spatial view cells activate when the primate gazes at their respective preferred domains in
		the visual field (left). The curves $r_1(t)$ and $r_2(t)$ traced by the monkey's gaze induce simplicial
		paths $\Gamma_1$ and $\Gamma_2$ running through the corresponding coactivity complex $\mathcal{T}_{VC}$
		(right).}
	\label{fig:fields} 
\end{figure}

Specifically, according to the \textit{Leray criterion}, a complex $\Sigma$ representable in $D$ dimensions
should not contain non-contractible gaps, cavities or other topological defects in dimensionalities higher 
than $(D-1)$ \cite{Leray}. Formally, it is required that the homological groups of $\Sigma$ and hence its Betti
numbers should vanish in these dimensions, $b_{i\geq D}(\Sigma)=0$. Moreover, the Betti numbers of all the 
subcomplexes $\Sigma_x$ of $\Sigma$, induced by a fraction $x$ of its vertexes should
also vanish, $b_{i\geq D}(\Sigma_x)=0$. In the case of coactivity complexes, such subcomplexes $\mathcal{T}_x
\subseteq\mathcal{T}$ have a particularly transparent interpretation: they are the ones generated by $x\%$ of
the active cells. According to the second criterion, the number of simplexes in all dimensions of $\Sigma$ must
obey \textit{Eckhoff's inequalities}---a set of combinatorial relationships discussed in 
\cite{Eckhoff,Kalai1,Kalai2,Kalai3} and listed in the Sec.~\ref{sec:met}, where we also briefly detail the
Leray criterion \cite{TancerSur,Leray,Kalai3,KM1,KM2}. 

Previous topological studies of the coactivity data were motivated by the Alexandrov-\v{C}ech theorem
\cite{Alexandroff,Cech,Hatcher,Edwards}, which posits that the homologies of the nerve complexes produced by
the ``good" covers (i.e., the ones with contractible overlaps (\ref{overlap}), see \cite{Tancer3}), should 
match the homologies of the underlying space $X$, $H_{\ast}(\mathcal{N})= H_{\ast}(X)$, i.e., have the same
number of pieces, holes, tunnels, etc. Specifically, this construction was applied to the place cell coactivity
complexes, whose representability was presumed \cite{Ghrist1,Curto,PLoS}. Persistent homology theory 
\cite{Ghrist,Kang,Wasserman,Zomorodian,EdelZom,ZomorodianBook} was used to trace the dynamics of the Betti
numbers $b_{i \leq D}(\mathcal{T})$ in physical dimensionalities $D\leq2$ \cite{Arai,Basso,CAs,MWind2,Eff,Rev}
and $D\leq3$ \cite{Hoffman,Rev}, to detect whether and when these numbers match the Betti numbers of the 
environment, $b_{i\leq D}(\mathcal{E})$, and how this dynamics depends on spiking parameters. 
It was demonstrated, e.g., that for a wide range of the firing rates and place field sizes referred to as the
\textit{Learning Region}, $\mathcal{L}(\mathcal{E})$, the low-dimensional Betti numbers of $\mathcal{T}(t)$
converge to their physical values after a certain period $T_{\min}$, neurobiologically interpreted as the
minimal time required to ``learn" the  topology of the environment ($b_{i\leq D}(\mathcal{T}(t))= b_{i\leq D}
(\mathcal{E})$, $t\geq T_{\min}$ \cite{PLoS}). 

Moreover, it became possible to asses the contribution of various physiological parameters---from brain waves
to synapses---to producing and sustaining the topological shape of $\mathcal{T}$ \cite{Arai,Basso,Hoffman,CAs,MWind2,Eff,Rev}.
In addition, the coactivity complexes were used for contextualizing the ongoing spiking activity and linking
its structure to the animal's behavior. For example, it was shown that a trajectory $\gamma(t)$ tracing through
a sequence of firing domains $\upsilon_{\sigma_i}$, produces a ``simplicial path" $\Gamma$---a succession of 
active simplexes that captures the shape of $\gamma(t)$ and allows interpreting the animal's active behavior 
\cite{Brown1,Barbieri1,Jensen1,Guger} and its ``off-line" memory explorations \cite{Karlsson1,Wu,Ji,Johnson,
	Dragoi,Pfeiffer1,TaubeGood,Buffalo} (Fig.~\ref{fig:fields}).

Together, these arguments suggest that experimentally discovered representing spaces and firing fields serve as
explicit models of the cognitive maps emerging from neuronal activity---a perspective that is currently widely
accepted in neuroscience. However, this view requires verification, since the empirically identified firing 
fields may be contextual offshoots or projections from some higher-dimensional constructs---in the words of H.
Eichenbaum, ``\textit{hippocampal representations are maps of cognition, not maps of physical space}" \cite{Viewpoints}. 
The way of addressing this question is straightforward: if the spiking activity is intrinsically spatial, i.e.,
if neurons represent spatial domains, then the coactivity complexes generated by the corresponding neuronal
ensembles should be representable---an explicit property that can be confirmed or refuted using Leray, Eckhoff
and other criteria. In the following, we apply these criteria to several types of neuronal activity, both 
simulated and experimentally recorded, and discuss the results. 

\section{Results}
\label{sec:res}

\textbf{Simplicial topology approach}. The conventional theory of representability addresses properties of 
``static" simplicial complexes \cite{TancerSur,Tancer2,Kratochvil,Matousek2,Tancer3,Leray,Fulkerson,Kratsch,
Habib,Golumbic,Fishburn,Eckhoff,Kalai1,Kalai2,Kalai3,KM1,KM2}.
In contrast, the coactivity complexes are dynamic structures that can be viewed as time-ordered agglomerates 
of simplexes, restructuring at the moments $t_1<t_2<t_3\ldots$, 
\begin{equation}
\mathcal{T}(t_1)\subseteq \mathcal{T}(t_2)\subseteq \mathcal{T}(t_3)\ldots\,.
\label{filt}
\end{equation}
The exact organization of each complex in the sequence (\ref{filt}) depends on the specifics of the underlying
spiking activity, e.g., the initial state of the network, its subsequent dynamics, spiking mechanisms and so 
forth (in case of the place fields, think of the starting point of navigation, shape of the trajectory, speed,
etc.). Thus, verifying representability of these complexes requires testing whether Eckhoff, Leray and other
criteria are valid at each moment $t$.

We constructed coactivity complexes by simulating the rat's navigation through a planar environment $\mathcal{E}$
commonly used in electrophysiological experiments (Fig.~\ref{fig:fields}A, see also \cite{Hafting,BrunG}). The
neuronal spikings in this case are generated as responses to the rat's appearances within preconstructed, convex
firing domains, e.g., stepping into randomly scattered place fields or facing towards head direction fields 
centered around randomly chosen preferred angles (see Sec.~\ref{sec:met} and Methods in \cite{PLoS,Arai,CogAff}).
While the resulting nerve complexes (\ref{nerve}) are $2D$-representable by design, we inquired whether the 
corresponding coactivity complexes are also representable, i.e., whether the activity of individual neurons 
intrinsically represents regions and whether connectivity between these regions is similar to the connectivity
between the underlying auxiliary firing fields.

\begin{figure}[h]
	\includegraphics[scale=0.8]{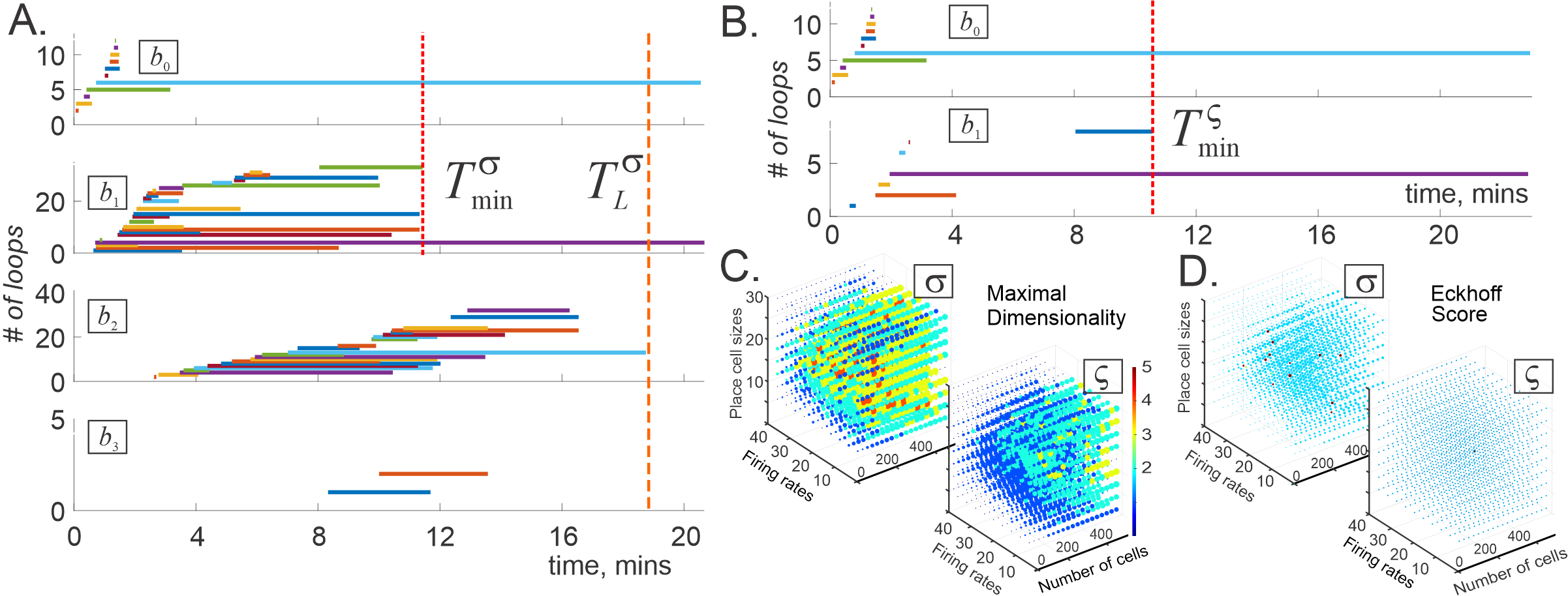}
	\caption{\footnotesize \textbf{Persistent Leray dimension}.
		(\textbf{A}). The Leray dimensionality of the coincidence-detector complex $\mathcal{T}_{\sigma}(t)$
		constructed for an ensemble of $N_c=300$ place cells can rise to $D(\mathcal{T}_{\sigma})=4$ (here the
		mean maximal firing rate is $f=12$ Hz, mean place field size $s=22$ cm; environment $\mathcal{E}$ same
		as on Fig.~\ref{fig:fields}A). In about $17$ minutes---the corresponding Leray period $T_L$---the 
		dimensionality drops to $D(\mathcal{T}_{\sigma})=\bar{D}_L(\mathcal{T}_{\sigma})=1$, after which the
		spiking patterns can be intrinsically interpreted in terms of planar firing fields. Note that the Leray
		period in this case is longer than the minimal learning time evaluated based on the lower-dimensional
		Betti numbers $b_{0,1}(\mathcal{T}_{\sigma})$, $T_L > T_{\min}^{\sigma}$. Shown are all the non-zero
		Betti numbers of $\mathcal{T}_{\sigma}(t)$.
		(\textbf{B}). Timelines of the topological loops in a spike-integrating coactivity complex, evaluated
		for the same cell population in the same environment yields the persistent Leray dimensionality 
		$\bar{D}_L(\mathcal{T}_{\varsigma})=1$ from the onset. The disappearance of spurious $0D$ loops in about
		$11$ minutes marks the end of the learning period $T_{\min}^{\varsigma}(t)$. Note that the number of
		spurious loops in $\mathcal{T}_{\varsigma}(t)$ is significantly lower than in $\mathcal{T}_{\sigma}(t)$.
		\textbf{C}. Maximal dimensionality of the topological loops in $\mathcal{T}_{\sigma,\varsigma}(t)$.
		(\textbf{D}). The Eckhoff conditions are satisfied for nearly all coincidence-detecting complexes
		$\mathcal{T}_{\sigma}(t)$ (left panel, occasional exceptions are shown by red dots) and for all spike
		integrating complexes $\mathcal{T}_{\varsigma}(t)$ (right panel).
	}
	\label{fig:eckler}
\end{figure}

Simulations show that \textit{persistent Leray dimensionality} $\bar{D}_L$ (above which the spurious loops in
$\mathcal{T}(t)$ vanish, $\bar{D}_L=\min(\{D:b_{i>D}(\mathcal{T}(t))=0,\,t\geq T_L\})$, see also \cite{Curto1})
eventually settles at $\bar{D}_L =1$ for most complexes, implying that neuronal activity defines a proper planar
map. However, this mapping requires time---a \textit{Leray period} $T_L$---which, for the maps populating the
learning region $\mathcal{L}(\mathcal{E})$, is typically similar to the learning time $T_{\min}$ 
(Fig.~\ref{fig:eckler}A,B).

Whether a particular value of $T_L$ is shorter or longer than the corresponding $T_{\min}$ depends on how 
exactly the coactivity complex is constructed, e.g., whether the simplexes (\ref{sigma}) correspond to 
simultaneously igniting cells groups or assembled from lower-order combinations over an extended period 
$\varpi$ \cite{Syntax}. Physiologically, the former corresponds to the case when spiking outputs are processed
by ``coincidence-detector" neurons in the downstream networks and the latter to the case when lower-order
coactivities are collected over a certain ``spike integration window" $\varpi$---longer than the simultaneity
detection timescale $w$ \cite{Konig,London,Koulakov}. Different readout neurons or networks may have different
integration periods; to simplify the model, we started by extending the parameter $\varpi$ to the entire 
navigation period for all cells and cell groups.

The lowest order of coactivity involves spiking cell pairs \cite{Spruston}, which together define a coactivity
graph $\mathcal{G}$ \cite{Muller,Burgess}. The cliques $\varsigma$ of this graph produce a \textit{clique
coactivity complex} $\mathcal{T}_{\varsigma}$ that generalizes the \textit{simplicial coactivity complex}
$\mathcal{T}_{\sigma}$, built from simultaneously detected simplexes \cite{Basso,Hoffman,CAs,MWind2,Eff,Rev}.
As it turns out, the ``coincidence detecting" and the ``spike integrating" complexes have different topological
dynamics: the former are more likely to start off with a higher Leray dimensionality, $D_L\geq 3$, that
then reduces to $\bar{D}_L\leq 2$ (Fig.~\ref{fig:eckler}A), whereas the latter tend to be more stable, 
lower-dimensional and have shorter Leray and learning times (Fig.~\ref{fig:eckler}B).

To test the induced subcomplexes of each $\mathcal{T}$, we selected random subcollections of cells containing
$x=50\%$, $x=33\%$, $x=25\%$ and $x=20\%$ of the original neuronal ensemble, and found that if the original complex
$\mathcal{T}\equiv\mathcal{T}_{x=1}$ is representable, then its subcomplexes, $\mathcal{T}_{x<1}\subseteq\mathcal{T}$,
typically require less time to pass the Leray criterion, $T_L(\mathcal{T}_{x<1})\leq T_L(\mathcal{T})$, and that
for $x>50\%$ the Leray times saturate, $T_L(\mathcal{T}_{x>0.5})\approx T_L(\mathcal{T})$. Thus, the Leray time
of the full complex, $T_L(\mathcal{T})$, can be used as a general estimate of the timescale required to establish
representability.

To control the sizes of the coactivity complexes, we used only those periods of each neuron's activity when it
fired at least $m$ spikes per coactivity window ($w\approx 1/4$ secs; for justification of this value see 
\cite{Mizuseki,Arai}). Additionally, we used only those groups of coactive cells in which pairwise coactivity
exceeded a threshold $\mu$ (Sec.~\ref{sec:met}). Biologically, these selections correspond to using only the 
most robustly firing cells and cell assemblies for constructing the coactivity complexes \cite{CAs}. The results
demonstrate that majority of the coactivity complexes $\mathcal{T}$ computed for smallest
possible $m$ and $\mu$ exhibit low \textit{persistent Leray dimensionality}, $\bar{D}_L =1$, which points at
$2D$ representability of the underlying neuronal activity, with the Leray times $T_{L}$ 
similar to the corresponding learning times $T_{\min}$ (Fig.~\ref{fig:eckler}C).
We also found that Eckhoff inequalities are typically satisfied throughout the navigation period, i.e., that the
Eckhoff criterion does not significantly limit the scope of representable spiking in this case 
(Fig.~\ref{fig:eckler}D).

\textbf{2. Region Connection Calculus (\textsf{RCC})}. An independent perspective on spatial representability
is provided by Qualitative Space Representation approach (\textsf{QSR}, \cite{CohnRenz,ChenCohn}), which sheds
a new light on the dynamics of neuronal maps. From \textsf{QSR}'s perspective, a population of cells 
$\mathfrak{C}=\{c_1,c_2,\ldots,c_N\}$ may represent a set of abstract, or \textit{formal} spatial regions 
$R=\{r_1,r_2,\ldots,r_N\}$, if the relationships between them, as defined by the cells' coactivity, can be
consistently actualized in a topological space $X$ by a set of explicit regions, $\Upsilon=\{\upsilon_1,
\upsilon_2,\ldots\upsilon_N\}$.

Specifically, regions $r_i$ and $r_j$ encoded by the cells $c_i$ and $c_j$ can be:
\vspace{5 pt}
\begin{enumerate}[nosep]
\item disconnected, $\mathsf{DR}(r_i,r_j)$, if $c_i$ and $c_j$ never cofire;
\item equal, $\mathsf{EQ}(r_i,r_j)$, if $c_i$ and $c_j$ are always active and inactive together;
\item proper part of one another, if $c_j$ is active whenever $c_i$ is, $\mathsf{PP}(r_i,r_j)$, 
or vice versa, $\mathsf{PPi}(r_i,r_j)$;
\item partially overlapping, $\mathsf{PO}(r_i,r_j)$, if $c_i$ and $c_j$ are sometimes (but not always) coactive.
\end{enumerate}
\vspace{5 pt}

\begin{figure}[h]
	\vspace{-10pt}
	\includegraphics[scale=0.82]{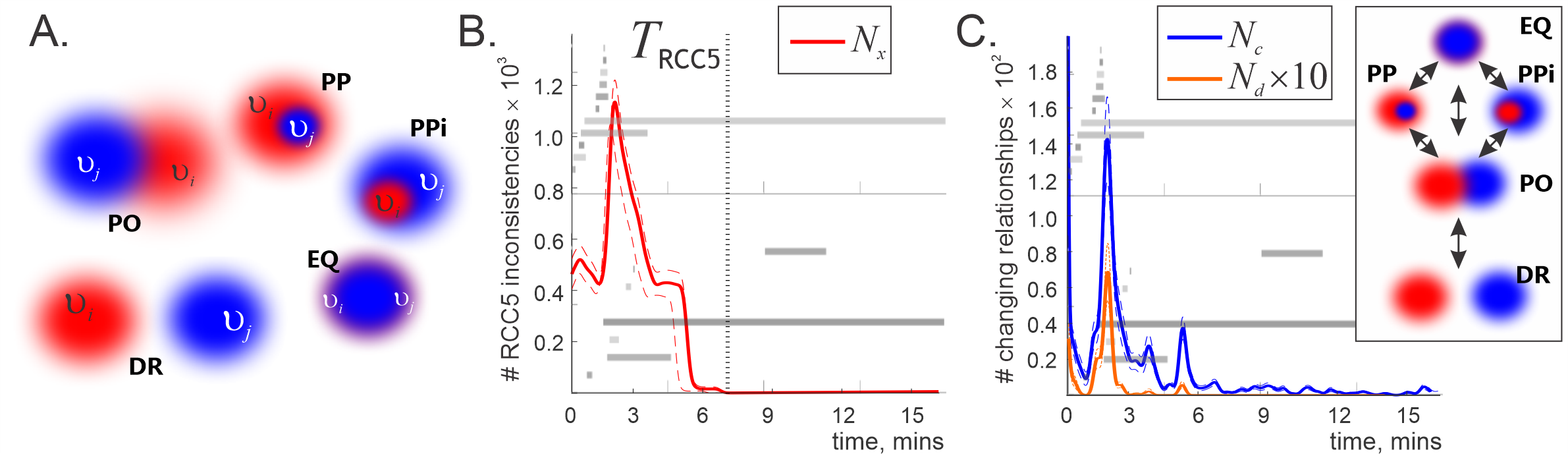}
	\caption{{\footnotesize\textbf{\textsf{RCC5} analyses}. 
			\textbf{A}. Two regions with soft boundaries, e.g. two firing fields $\upsilon_i$ and $\upsilon_j$, can
			overlap, $\textsf{PO}(\upsilon_i,\upsilon_j)$, be proper parts of each other, $\textsf{PP}(\upsilon_i,
			\upsilon_j)$ or $\textsf{PPi}(\upsilon_i,\upsilon_j)$, be disconnected $\textsf{DR}(\upsilon_i,\upsilon_j)$
			or coincide $\textsf{EQ}(\upsilon_i,\upsilon_j)$.
			\textbf{B}. Number $N_x(t)$ of inconsistent triples of \textsf{RCC5} relationships appearing in the
			$\mathcal{R}_5(t)$ relational framework constructed for the same neuronal ensemble as illustrated in
			Fig.~\ref{fig:eckler}. The barcode diagram for the corresponding integrating coactivity complex 
			(Fig.~\ref{fig:eckler}B) is shown in the background, to illustrate the correspondence between the
			\textsf{RCC5} and the homological dynamics. $T_{\R5}$ (dotted line) marks the time when inconsistencies
			in the $\mathcal{R}_5(t)$ schema disappear. Results averaged over $10$ repetitions, error margins shown
			by the dashed lines.
			\textbf{C}. The net number of changes of \textsf{RCC5}-relationships between two subsequent moments of
			time, $N_c(t)$, shown by the blue line, and the number $N_d(t)$ of changes that violate the \textsf{RCC5}
			continuity order (top right panel), shown by the orange line. For better illustration, $N_d(t)$ is scaled
			up by a factor of $10$.	Initially, discontinuous events are frequent but shortly before $T_{\R5}$ they
			disappear entirely, leaving the stage to qualitatively continuous sequences. The same barcodes are added 
			in the background, error margins shown by dashed lines.
	}}
	\label{fig:rcc}
\end{figure}

These five relations fully capture mereological configurations of regions in a first-order logical calculus
known as \textsf{RCC5} (Fig.~\ref{fig:rcc}A, \cite{Cohn94g}). Using mereological, rather than topological,
distinctions reflects softness of the firing fields' boundaries: the probabilistic nature of neuronal spiking
does not warrant determining whether two regions actually abut each other or not. 

A key property of a \textsf{RCC5}-framework defined by spiking neurons---a $\mathcal{R}_5$ schema---is its
internal consistency \cite{SchemaS}. It may turn out, e.g., that some pairs of cells encode relationships that
are impossible to reconcile, e.g., $\mathsf{PP}(r_i,r_j)$, $\mathsf{DR}(r_j,r_k)$ and $\mathsf{PO}(r_i,r_k)$.
Indeed, if an actual region $\upsilon_i$ is contained in $\upsilon_j$ then it cannot possibly overlap with a 
region $\upsilon_k$ that is disconnected from $\upsilon_j$. Correspondingly, the neuronal activity that produces
such inconsistencies (for the full list
see Table~\ref{tab:comp} in Sec.~\ref{sec:met}) is not representable---not even interpretable in spatial terms.
On the other hand, it can be shown that if all triples of relationships are consistent, then $\mathcal{R}_5(t)$
does possess a spatial model, i.e., there exists a set of regions $\upsilon_i$ (with no prespecified properties
such as convexity, connectivity or dimensionality) that relate to each other as the $r_i$s relate in 
$\mathcal{R}_5$ \cite{Renz,Cohn94g,CohnRenz,ChenCohn,Bennet1,Long}.

To verify whether spiking activity is representable in this \textsf{QSR} sense, we constructed an inflating
$\mathcal{R}_5(t)$-schema (an \textsf{RCC5}-framework growing as spiking data accumulates, similar to 
(\ref{complex})) for each neuronal ensemble and counted the inconsistent triples of relationships at 
each moment $t$. The results show that all $\mathcal{R}_5(t)$-schemas start off with numerous inconsistencies,
which tend to disappear after a certain period $T_{\R5}$ that is typically smaller than the Leray time $T_L$ 
(Fig.~\ref{fig:rcc}B,C). 

The net dynamics of \textsf{RCC5} relationships is illustrated on Fig.~\ref{fig:rcc}C. Note that some of these
changes may be attributed to the regions' continuous reshapings or displacements, e.g., two overlapping regions
may become disconnected, $\textsf{PO}(r_i,r_j)\to\textsf{DR}(r_i,r_j)$, is $r_i$ moves away from $r_j$, or $r_i$
may move into $r_j$, inducing $\textsf{PO}(r_i,r_j)\to\textsf{PP}(r_i,r_j)$. In contrast, a jump from a 
disconnect to a containment, without an intermediate partial overlap, e.g., $\textsf{DR}(r_i,r_j)\to\textsf{PP}
(r_i,r_j)$ rather than $\textsf{DR}(r_i,r_j) \to \textsf{PO}(r_i,r_j) \to\textsf{PP}(r_i,r_j)$, would be a 
discontinuous, abrupt change. As shown on Fig.~\ref{fig:rcc}C, discontinuous transitions are common at the 
initial stages of navigation, but shortly before $T_{\R5}$ they disappear, indicating that the relationships 
between regions encoded within a sufficiently well-developed $\mathcal{R}_5(t)$ schema evolve in a continuous 
manner.

These outcomes not only provide an alternative lower-bound estimate for the time required to accumulate data 
for producing low-dimensional spatial representations, but also help understanding the nature of processes 
taking place prior to Leray time. In particular, the exuberant initial dynamics, homologically manifested 
through an incipient outburst of spurious loops in the coactivity complexes (Figs.~\ref{fig:eckler}A,B, 
$t<T_{\min}$), cannot be interpreted as a mere ``settling" of topological fluctuations in the cognitive 
map---according to Fig.~\ref{fig:rcc}B, the $\R5$-schema does not form a coherent topological stratum for
$t<\mathcal{T}_{\R5}$. Rather, the initial disorderly period should be viewed as the time of transition from
a nonspatial to a spatial phase, followed by spatial dynamics (for $t>\mathcal{T}_{\R5}$) that involves, 
\textit{inter alia}, dimensionality reduction and other restructurings (Fig.~\ref{fig:rcc}C).

\textbf{3. Current summary}. Taken together, these results show that even in the simplest ``reactive" model, in
which neuronal firings are simulated as responses to regular domains covering a compact space, the low-dimensional
representability is not an inherent, but an emergent property. In particular, \textsf{RCC5}-analyses suggest
that spatial interpretation of neuronal spiking becomes possible after a finite period. During the times that
exceed both $T_{\R5}$ and $T_L$, the spiking data can be interpreted in terms of firing fields in a space $X$
of dimensionality higher than the persistent Leray dimensionality of the corresponding coactivity complex,
$\dim(X)>\bar{D}_L(\mathcal{T})$. 

Physiologically, this implies that the outputs of place cells, head direction cells, view cells, etc., may not
be immediately interpretable by the downstream networks as representations of spatial regions---the information
required for such inference appears only after a certain ``evidence integration." Correspondingly, the firing
field maps constructed according to the standard experimental procedures \cite{Dostrovsky,Taube,Rolls1} also
cannot be considered as automatic ``proxies" of cognitive maps: such interpretations are appropriate only after
the representability of the corresponding coactivity complexes is established. Another principal conclusion is
that representability of the spiking activity depends not only on the spiking outputs, but also on how the 
information carried by these spikes is detected and processed. In particular, spikes integrated over extended
periods are likelier to permit a consistent firing field interpretation than spikes counted via coactivity 
detection. On the technical side, these results imply that an accurate description of the firing fields' 
plasticity should include possible dimensionality changes  \cite{Barbieri,FrankPl1,Eden}. 

\begin{figure}[h]
	\includegraphics[scale=0.845]{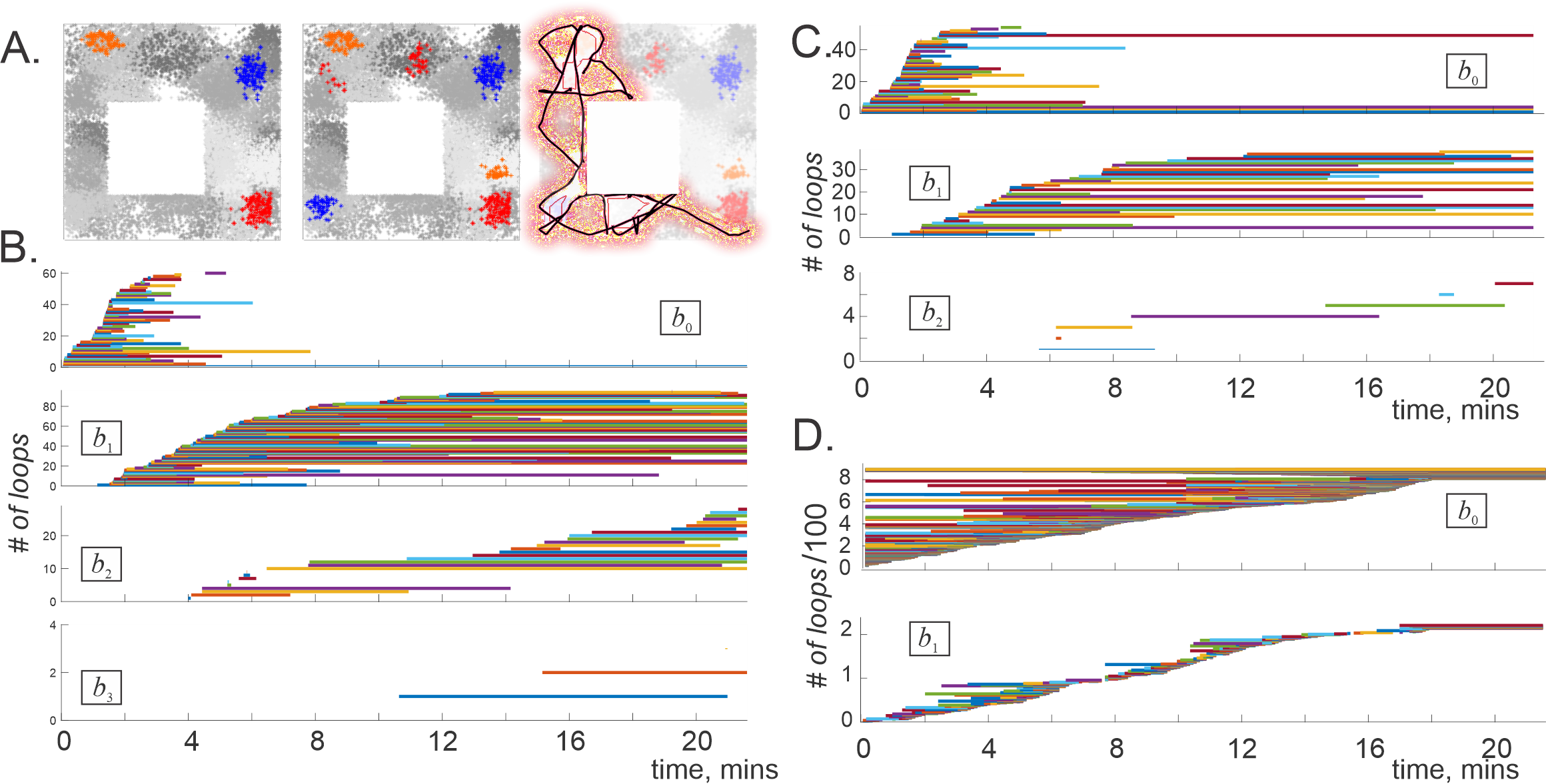}
	\caption{\footnotesize \textbf{Topological dynamics in maps with multiple firing fields}. (\textbf{A}).
		Left panel shows three examples of convex place fields used to obtain the results illustrated in
		Fig.~\ref{fig:eckler}. Allowing a cell to spike in several ($2-3$) locations produces multiply 
		connected place fields (middle panel; clusters of dots of a given color correspond to spikes produced
		by a single simulated neuron). Right panel shows a $\varpi=50$ second long fragment of the trajectory
		$\gamma_{\varpi}$ covering a segment $\chi_{\varpi}$ of the environment (reddened area). 
		(\textbf{B}). The Leray dimensionality of the detector-complex evaluated for the same place cell 
		population as in Fig.~\ref{fig:eckler}A, with the exception of $30\%$ of multiply connected place 
		fields ($2-3$ components each) can reach $D(\mathcal{T}_{\sigma})=4$.
		(\textbf{C}). In a clique coactivity complex, the spurious loops in dimensions $D=2$ and lower may 
		persist indefinitely, implying either that the firing fields are $3D$-representable \textit{or} that
		they may be multiply connected. Note that the number of spurious loops in both $\mathcal{T}_{\sigma}$
		and in $\mathcal{T}_{\varsigma}$ is higher than in the case with convex firing fields 
		(Fig.~\ref{fig:eckler}A,B).
		(\textbf{D}). The persistence bars computed for the flickering complex $\mathcal{F}_{\varpi}$ with 
		spike integration window $\varpi=1$ minute, indicate stable mean Leray dimensionality 
		$\langle D(\mathcal{F}_{\varpi})\rangle=1$,
		implying that the local charts $\chi_{\varpi}$ are planar and hence that the firing fields are two-dimensional.
	}
	\label{fig:mpfs}
\end{figure}

\textbf{4. Multiply connected place fields}. A key simplification used in the simulations described above is
that firing fields were modeled as convex regions. While this assumption is valid in some cases \cite{Brown2},
multiply connected firing fields are also commonly observed (Fig.~\ref{fig:mpfs}A, \cite{Singer,KnierimInt}).
From our current perspective, the issue is that multiple connectivity of the cover elements (\ref{cover}) may
increase the Leray dimensionality of the corresponding nerve complex \cite{TancerSur,Amenta} and thus bring
additional ambiguity into the analyses. Identification of the firing fields' connectivity from the spiking data
is an elaborate task that requires tedious analyses of the spike trains produced by individual cells or cell
groups over periods comparable to the Leray and the learning times \cite{Curto1,Curto2}. To circumvent these
difficulties, we reasoned as follows.

Suppose that the spiking activity used to produce a coactivity complex $\mathcal{T}(t)$ is generated as a
moving agent (animal's body, its head, its gaze) follows a trajectory $\gamma(t)$ over a space $X$, covered 
with stable firing fields $\upsilon_i$, $i=1,\ldots,N$. Consider a navigation period $\varpi$ that spans over
a smaller segment of this trajectory, $\gamma_{\varpi}=\{\gamma(t): t\in\varpi\}$. If $\varpi$ is sufficiently
short, then one would expect $\gamma_{\varpi}$ to cross at most one component of a typical firing field 
$\upsilon_i$; even if $\gamma_{\varpi}$ meets more than one component of a multiply connected $\upsilon_i$, 
this property may not manifest itself in the resulting spike trains, i.e., $\upsilon_i$ should be 
\textit{effectively} simply connected (Fig.~\ref{fig:mpfs}A). Correspondingly, the Leray dimensionality of 
the coactivity complex acquired during that period should reflect the dimensionality of a small underlying 
fragment of $X$---a \textit{local chart} $\chi_{\varpi}$---that contains $\gamma_{\varpi}$ (topologically, 
$\chi_{\varpi}(\gamma)\cong\{\cup_j\upsilon_j|\gamma_{\varpi} \cap\upsilon_j\neq\varnothing\}$). The 
dimensionality of $\chi_{\varpi}$ can then be ascribed to all the contributing $\upsilon_j$s, $\dim(\upsilon_j)
=\dim(\chi_{\varpi})$.

Further, if the $\varpi$-period is allowed to shift in time, then the segment $\gamma_{\varpi}$ will also slide
along the trajectory $\gamma(t)$; the spikes fired within each $t$-centered window, $\varpi_t = [t-\varpi/2,t+
\varpi/2]$, will then produce a $\varpi_t$-specific \textit{flickering coactivity complex} $\mathcal{F}_{\varpi}
(t)\subseteq \mathcal{T}(t)$, whose topological properties may change with time \cite{MWind2,PLoZ,Replays}.
Since $\mathcal{F}_{\varpi}(t)$ contains a finite number of elements, it will reconfigure at discrete moments,
$t_1,t_2,\ldots$, and remain unchanged in-between, $\mathcal{F}_{\varpi}(t)=\mathcal{F}_{\varpi}(t_k)$, $t \in
[t_k,t_{k+1})$. If a given instantaneous configuration $\mathcal{F}_{\varpi}(t_k)$ is representable, then its 
vertexes correspond to the regions comprising the local chart $\chi_{\varpi}(t_k)$, with dimensionality $\dim(
\chi_{\varpi}(t_k))\geq D_L(\mathcal{F}_{\varpi}(t_k))$. If two such complexes overlap, $\mathcal{F}_{\varpi}
(t_k)\cap\mathcal{F}_{\varpi}(t_{l})\neq\varnothing$ (i.e., their vertex sets overlap), then their respective
charts also overlap $\chi_{\varpi}(t_k)\cap\chi_{\varpi}(t_{k+1})\neq\varnothing$, which allows relating their
topological properties, including properties of the representing regions. 

Clearly, the outcome may depend on how each $\gamma_{\varpi}$ is embedded into $X$, the spiking parameters, etc.
Moreover, since the Leray dimensionality of the instantaneous complexes can change, so can the dimensionalities
of the corresponding local charts: $D_L(\mathcal{F}_{\varpi}(t_k))\neq D_L(\mathcal{F}_{\varpi}(t_l))$ may 
entail $\dim(\chi_{\varpi}(t_k))\neq \dim(\chi_{\varpi}(t_l))$. This may seem as a contradiction since the 
representing space is naturally assumed to be a topological manifold, i.e., all of its local charts, arbitrarily
selected, should have the same dimensionality $\dim(\chi_{\varpi}(t))=\dim(X)=D$. On the other hand, the 
deviations of the local dimensionality estimates from a fixed $D$ can be viewed as mere fluctuations caused by
occasional contribution of multiply connected firing fields or by other noise sources, e.g., by stochasticity
of neuronal spiking \cite{FentonVar}. One can hence attempt to discover the true dimensionality of $X$ by 
evaluating the mean Leray dimensionality of the instantaneous complexes, 
$$\dim(X)=\langle D_L(\mathcal{F}_{\varpi}(t_k))\rangle_k,$$ which physiologically alludes to learning the 
physical structure of the underlying space from the recurrent information.

Numerical verification of the viability of the proposed approaches can be achieved by simulating multiply
connected firing fields and computing homological dynamics of the resulting coactivity complexes. To that end,
we randomly added $2-3$ additional convex components to $\sim 30\%$ of the place fields (Fig.~\ref{fig:mpfs}A)
and simulated the topological dynamics of the corresponding complexes.

The results show that multiple connectivity of the firing fields does indeed increase Leray dimensionality in
both the detector and the integrator complexes, $\mathcal{T}_{\sigma}(t)$ and $\mathcal{T}_{\varsigma}(t)$.
Moreover, in contrast with the complexes generated off the maps with convex fields, the maps with multiply
connected fields tend to produce persistent higher-dimensional loops, notably in the coactivity detecting 
complexes $\mathcal{T}_{\sigma}(t)$ (compare Fig~\ref{fig:eckler}A and Fig.~\ref{fig:mpfs}B). In the spike
integrating clique complex $\mathcal{T}_{\varsigma}(t)$, the Leray dimensionality remains low and may in some
cases retain the physical value $D_{L}(\mathcal{T}_{\varsigma})=1$, although topological loops in dimensions
$D=2$ and even higher may also appear (Fig.~\ref{fig:mpfs}C). Thus, multiple firing field connectivity 
significantly increases the number of spurious $1D$ holes (by $200-300\%$), precluding both types of complexes
from assuming the physically expected topological shapes. 

Tighter dimensionality estimates can be produced by using shorter spike integration windows $\varpi\lesssim
T_{\min}$ and constructing flickering coactivity complexes $\mathcal{F}_{\varpi}(t)$ from pairwise coactivities
detected over $\varpi$-periods shifting by discrete steps $\Delta\varpi$ and yielding an array of windows 
$\varpi_1,\varpi_2,\varpi_3,\ldots$ centered at $t_k =\varpi/2+(k-1)\Delta\varpi$. The specific $\varpi$-values
were chosen comparable to the characteristic time required by the rat to run through a small segment of the 
environment: $\varpi\approx 25-65$ secs for the arena shown on Figs.~\ref{fig:fields}A and \ref{fig:mpfs}A.
The Betti numbers for this case were evaluated using zigzag homology theory---a generalization of the 
persistent homology theory that applies to complexes that can not only grow, but also shrink, break apart, 
fuse back again, etc. \cite{Carlsson1,Carlsson2}. In particular, this approach allows studying how the 
topological fluctuations in $\mathcal{F}_{\varpi}(t)$ affect its Leray dimensionality 
$D_L(\mathcal{F}_{\varpi}(t))$ from moment to moment.

Typical results illustrated on Fig.~\ref{fig:mpfs}D show that there appears a large number of spurious $0D$
loops---disconnected pieces---with lifetimes nearly exponentially distributed about the learning periods 
$T_{\min}^{\varsigma}$, which suggests that fragments of $\mathcal{F}_{\varpi}(t)$ appear and disappear at 
random over such periods. The transient $1D$ loops also form and decay at $\varpi$-timescale. However, the most
important outcome is that the topological dynamics in dimensions $D>1$ trivializes---the higher dimensional 
loops in $\mathcal{F}_{\varpi}$ occur very rarely, if ever. These properties are qualitatively unaffected by
varying the discretization step $\Delta\varpi$ ($\varpi/20\lesssim\Delta\varpi\lesssim\varpi/10$) or changing
the window width $\varpi$, i.e., the estimates of the mean Leray dimensionality 
$\langle D_L(\mathcal{F}_{\varpi})\rangle=1$ are stable and reveal physical planarity of the representing space.

Verification of the \textsf{RCC5}-consistency of the spiking data produces the same qualitative results as in 
the case with simply connected firing domains: the $\mathcal{R}_5(t)$-schemas become consistent after a learning
period $T_{\R5}< T_{\min}^{\varsigma}$, upon which neuronal activity becomes spatially interpretable, and, by 
the Leray and Eckhoff arguments, representable in dimensions $D\geq \langle D_L(\mathcal{F}_{\varpi}) \rangle$.

\begin{figure}[h]
	\includegraphics[scale=0.8]{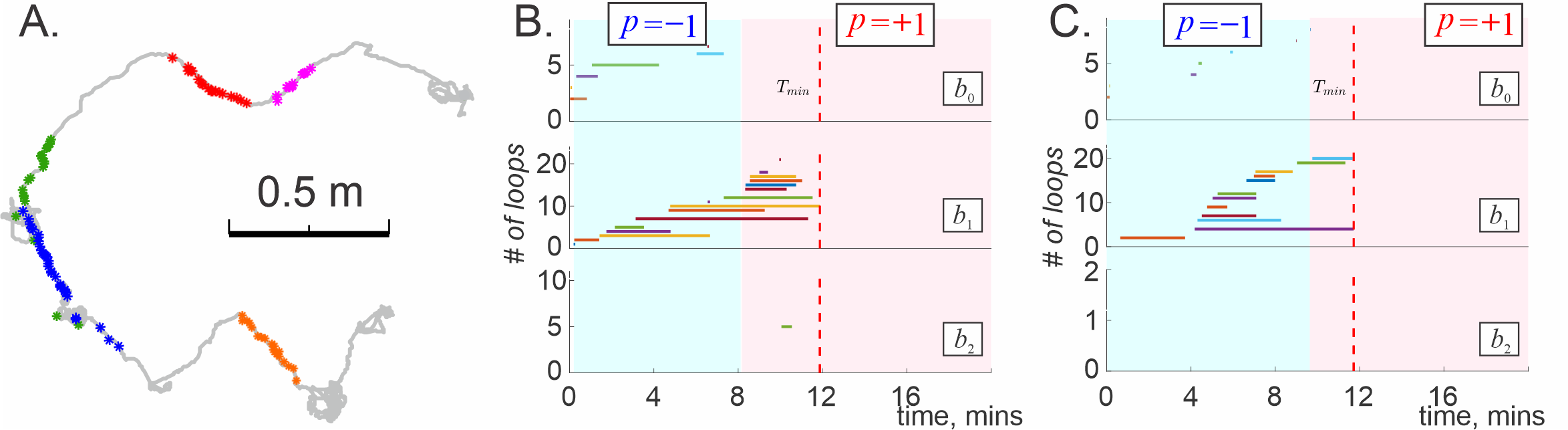}
	\caption{\footnotesize \textbf{Multiple firing fields}.
		(\textbf{A}) Spikes produced by five place cells (dots of different color) recorded in hippocampal CA1
		area of a rat navigating a linear track (speed $v\geq 3$ cm/sec). The underlying gray line shows a 
		fragment of the rat's trajectory (for more details see \cite{eLife}). 
		(\textbf{B}) Spurious topological loops in the corresponding coactivity complex disappear in $T_L\approx
		12$ minutes, revealing persistent Leray dimensionalities $\bar{D}=0$. The blue background highlights 
		the period during which the coactivity complex computed using only cells with convex place fields is
		not $1D$ representable ($p=-1$). The transition to $p(t)=+1$, marking the onset of $1D$ representability
		occurs at a time close to $T_L$.
		(\textbf{C}) Topological dynamics of the coactivity complex constructed using the data recorded during 
		the outbound moves only shows qualitatively similar behavior.
	} 
	\label{fig:dmpfs} 
\end{figure}

\textbf{5. Electrophysiological data}. We applied the analyses described above to spiking activity recorded
in the hippocampus (CA1 area) of rats navigating a linear environment shown on Fig.~\ref{fig:dmpfs}A (for more
data description and experimental specifications see \cite{eLife}). A typical running session, during which the
animal performed $45-70$ laps between the tips of the track, provided $N_c\lesssim25$ simultaneously recorded
neurons, allowing to construct small coactivity complexes that quickly become \textsf{RCC5}-consistent, comply
with the Eckhoff conditions, and exhibit persistent Leray dimensionality, $\bar{D}_L=0$, with typical persistent
Leray time $T_L\approx 10$ mins (Fig.\ref{fig:dmpfs}B). 

The vanishing $\bar{D}_L$ indicates that a linear track illustrated in Fig.~\ref{fig:dmpfs}A is contractible and
implies $1D$-representability. The latter can also be tested independently via \textsf{RCC5} analyses, which in 
this case allows identifying the track's linear structure \cite{MapSig}.

Since some of the hippocampal place fields are multiply connected, we also applied sliding window analyses, 
adjusting the spike integration period $\varpi_i$ to match the duration of the animal's $i^{\textrm{th}}$
run from one end of the track to the other (typically $2\leq \varpi_i \leq 10$ secs). Computations reveal
that the resulting complexes $\mathcal{F}_{\varpi_i}(t)$ exhibit the same mean Leray dimensionality 
$\langle D_L(\mathcal{F})\rangle =0$, which is consistent with the persistent $\bar{D}_L$ estimates above.
Combining these results produces convergent evidence that in this case the hippocampus does indeed map out
a $1D$ spatial domain (rather than $2D$, see \cite{eLife}).

The latter conclusion can, in fact, be verified by yet another representability test, which applies only to
$1D$ cases and presumes firing field convexity. The Golumbic-Fishburn (GF) algorithm \cite{Fulkerson,Kratsch,
Habib,Golumbic,Fishburn} is based on computing a binary index $p$: the $1D$-representable simplicial complexes
$\Sigma$ yield $p(\Sigma)=+1$, and the non-representable complexes produce $p(\Sigma)=-1$. For the inflating or
flickering coactivity complexes this index becomes time-dependent, $p=p(t)$, marking the evolution of $1D$ 
representability (Sec.~\ref{sec:met}).
Applying the GF-algorithm to the inflating coactivity complexes constructed for cells with convex place 
fields only, we found that $1D$ representability, $p(t)=+1$, appears in about $T_{+}\approx 10$ mins, close to
the Leray time (Fig.~\ref{fig:dmpfs}B,C), demonstrating consistency with the previously obtained results. 

Lastly, we addressed a particular property of the place cell's spiking activity in linear environments---the
place fields' directionality: a given place cell may fire during the outbound, but not inbound directions, or
vice versa \cite{BattDir}. We verified that the topological dynamics exhibited by the coactivity complexes built
using only the outbound or only the inbound activity are very similar to the dynamics of the full (bidirectional) 
complex $\mathcal{T}_{\varsigma}(t)$ (Fig~\ref{fig:dmpfs}B,C), implying that place cell directionality does
not necessarily compromise $1D$ representability of spiking activity. 

\section{Discussion}
\label{sec:disc}

Topological analyses of the spiking data allow testing whether a given type neuronal activity may arise from 
a ``spatial map," i.e., whether each neuron's spiking marks a domain similar to a place field, a head direction
field, a view field, etc., in a certain low-dimensional space. Thus far, establishing correspondences between
neurons and firing fields was based on matching the spike trains with spatial domains empirically, through 
trial and error \cite{Dostrovsky,Taube,Rolls1}. Here we attempt to address this question in a principled way,
through intrinsic analyses of the spiking data, without presuming or referencing \textit{ad hoc} constructions.
A set of hands-off algorithms discussed above allows objective estimates for the dimensionality of a space
needed to model the patterns of neuronal firing---a method that is unaffected by technical limitations, 
experimental ingenuity or complexity (e.g., nonlinearity) of the required firing field arrangements.

To follow the dynamics of the coactivity complexes we extend the conventional approaches of representability
theory into the temporal domain, obtaining several complementary time-dependent markers of representability. 
In particular, we use persistent homologies to extend Leray's theory to the case of inflating simplicial 
complexes and zigzag homologies in the case of flickering simplicial complexes. The latter approach is 
especially valuable as it allows extracting stable topological information from spiking data that may be 
generated from the maps with multiply connected firing fields or encumbered by other inherent irregularities,
in spirit with the general ideas of topological persistence \cite{Ghrist,Kang,Wasserman,Zomorodian,EdelZom,ZomorodianBook}.
It should also be mentioned that mathematical discussions of the persistent nerve theorem, alternative to 
ours and more formal, have began to appear \cite{Chazal,Cavanna}; however at this point our studies are 
independent.

A principal observation suggested by our analyses is that representability is a dynamic, emergent property that
characterizes current information supplied by the neuronal activity. Moreover, representability depends not 
only on the amount and the quality of the spiking data itself, but also on the mechanisms used for processing 
and interpreting this data. Both aspects affect the time required to establish the existence of a representing
space and its dimensionality. An implication of this observation is that experimentally constructed firing field
maps (place field maps, head direction maps, etc.) cannot be automatically regarded as direct models of cognitive
representations of ambient spaces \cite{Grieves,Taube,TaubeGood,Wiener,MosMc,Kropff,Derdikman} or more general
spatial frameworks \cite{Eichenbaum}; correctness of such interpretations may require more nuanced considerations.

\vspace{12pt}
\textbf{Acknowledgments}. The authors would like to thank Dr. M. Tancer for fruitful discussions and valuable
feedback and to D. Morozov for providing computational software. 
The work was supported by the European Research Council (ERC) under the European Unions Horizon 2020 Research
and Innovation program, grant 692854 (D.A.), by Alan Turing Institute Fellowship and EPSRC under grant EP/R031193/1
(A.C.) and by NSF grant 1901338 (Y.D.)

\newpage
\section{Methods}
\label{sec:met}

\subsection{Physiological parameters and constructions}
\vspace{3 pt}
$\bullet$ \textit{Simulated trajectory} $r(t)=(x(t),y(t))$, used for generating coactivity complexes was 
obtained by modeling a rat's non-preferential exploratory behavior---navigation without favoring of one
segment of the environment $\mathcal{E}$ over another (Fig.~\ref{fig:fields}A). The mean speed of about 
$\sim 20$ cm/sec was selected to match experimentally recorded speeds. The direction of the velocity 
$v(t)=(v_x(t),v_y(t))$ defines the ``angular trajectory" $\varphi(t)=\arctan{v_y(t)/v_x(t)}$ that traverses
the space of directions, $S^1$, allowing to simulate head direction cell activity as the rat explores 
$\mathcal{E}$ \cite{PLoS,Arai,CogAff}. The simulated \textit{navigation period}, $T = 25$ minutes, was
selected to match the duration of a typical ``running session" in electrophysiological experiments 
\cite{Brown2}. A shorter \textit{spike integration window} $\varpi\ll T$ was used to limit the pool of
spiking data for time-localized computations.

\vspace{3 pt}
$\bullet$ \textit{Poisson spiking rate} of a place cell $p$ depends on the animal's location $r(t)$, 
$$\lambda_p(r)=f_{p}e^{-\frac{|r-r_p|^2}{2s^2_{p}}},$$ where $f_p$ is the cell's maximal firing rate and $s_p$
defines the size of its place field \cite{Barbieri}. A similar formula defines the firing rate of a head direction
cell $h$, $\lambda_h(\varphi)$, as a function of the animal's ongoing orientation $\varphi$, the cell's preferred 
orientation angle $\varphi_h$, its maximal rate $f_h$ and the size of its preferred angular domain $s_h$. In
all simulations the firing fields were stable, i.e., the parameters of $\lambda_c$ and $\lambda_h$ remained 
constant.

\vspace{3 pt}
$\bullet$ \textit{Neuronal ensembles} produce lognormal distributions of the maximal firing rate amplitudes,
$f_c$, and of the firing field sizes, $s_c$ \cite{PLoS,BuzsakiLog}. We tested about $17,000$ different ensembles,
in which the ensemble mean maximal rate $f$ ranged between $4$ and $40$ Hz for the place cells and between $5$
to $35$ Hz for the head direction cells. The ensemble mean firing field sizes varied between $10$ to $90$ cm for
the place fields and between $12^{\circ} $ and $36^{\circ}$ degrees for the angular fields. For all ensembles, 
the firing field centers were randomly scattered over their respective representing spaces.

\vspace{3 pt}
$\bullet$ \textit{Multiple Firing Fields} were generated by adding two or three randomly scattered auxiliary
spiking centers $r_{c'}$, $r_{c''}$, etc., 
$$\lambda_c(r)=f_{c}e^{-\frac{|r-r_{c}|^2}{s^2_{c}}}+f_{c'}e^{-\frac{|r-r_{c'}|^2}{s^2_{c'}}}+\ldots\,\,\, .$$ 
The maximal firing rates at the auxiliary locations are smaller than the rate at the
main location, $f_c>f_{c'}>\ldots$, as suggested by the experiments \cite{Singer,KnierimInt}.

\vspace{3 pt}
$\bullet$ \textit{The activity vector} of a cell, $m_{c}=[m_{c,1},\ldots, m_{c,n}],$ is constructed by binning
its spike trains into $w=1/4$ seconds long ``coactivity windows" \cite{Arai,Mizuseki}. Each $m_{c,k}$ specifies
how many spikes were fired by $c$ into the $k^{th}$ time bin, $n$ is defined by the duration of navigation, 
$n=\floor{T/w}$. High activity periods can be identified by selecting time bins in which the number of fired
spikes exceeds an activity threshold $m$. 

\vspace{3 pt}
$\bullet$ \textit{Coactivity}. Two cells, $c_i$ and $c_j$, are \textit{coactive} over a time period $T$, if the 
formal dot product of their activity vectors does not vanish, $m_{ij}(T)=m_{c_i}(T) \cdot m_{c_j} (T)\neq 0$.
The set of all pairwise coactivities forms the coactivity matrix $M(T)=\|m_{ij}(T)\|$. Highly coactive pairs
of cells are the ones whose coactivity exceeds a threshold $\mu$. 


\subsection{Topological propaedeutics}
\textbf{Graphs}
\vspace{7 pt}

$\bullet$ \textit{A graph} $G$ is defined by its vertices, $V=\{v_1,v_2,\ldots,v_n\}$, and a set of edges $E$
that link certain pairs of vertexes. A formal description of a graph is given by its connectivity matrix $C(G)$,
with the elements
\begin{equation}
C_{ij}(G)= \begin{cases}
1, \,\,\, & \mbox{if $v_i$ and $v_j$ are connected by edge $e_{ij}$,} \\ 
0 \,\,\, & \mbox{if $v_i$ and $v_j$ are disconnected}.
\end{cases}
\nonumber
\label{connmtx}
\end{equation} 

\vspace{3 pt}
$\bullet$ \textit{A coactivity Graph} $\mathcal{G}$ is built by establishing functional links between cells
that exhibit high activity and coactivity ($m_{c,k}\geq m$, $m_{ij}\geq \mu$ see above) \cite{Burgess,Muller}.

\begin{figure}[h]
	\includegraphics[scale=0.8]{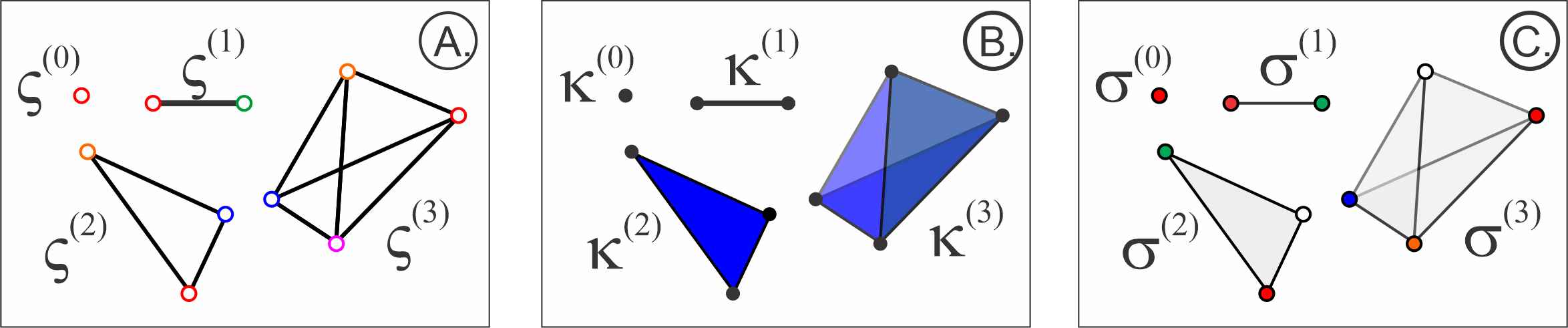}
	\caption{\footnotesize \textbf{Cliques and simplexes}.
		(\textbf{A}). Pairwise interlinked subsets of vertexes in graph $G$ form its cliques. Shown is a
		vertex $\varsigma^{(0)}$ ($0$-clique), a link $\varsigma^{(1)}$ ($1$-clique), a three-vertex 
		$\varsigma^{(2)}$ and a four-vertex $\varsigma^{(3)}$ cliques. 
		(\textbf{B}). Geometric simplexes: a $0D$ dot ($\kappa^{(0)}$), a $1D$ link ($\kappa^{(1)}$), a $2D$
		triangle ($\kappa^{(2)}$) and a $3D$ tetrahedron ($\kappa^{(3)}$).
		(\textbf{C}). The corresponding abstract simplexes: $\sigma^{(0)}$ (vertexes), $\sigma^{(1)}$ (pairs
		of vertexes), $\sigma^{(2)}$ (triples) and $\sigma^{(3)}$ (quadruples).
	} 
	\label{fig:simplx}
\end{figure}

\vspace{3 pt}
$\bullet$ A \textit{clique} of order $d$, $\varsigma^{(d)}$ in a graph is a fully interconnected subset of $(d+1)$
vertexes $v_{i_0},v_{i_1},\ldots,v_{i_d}$ (Fig.~\ref{fig:simplx}A).

\vspace{3 pt}
$\bullet$ Given a graph $G$, its \textit{complement graph} $\tilde{G}$ is produced by flipping $0$s and $1$s in
the connectivity matrix $C(G)$, i.e., joining the disconnected vertexes of $G$ and removing the existing edges. 

\vspace{3 pt}
$\bullet$ A \textit{comparability graph} $G_{\lhd}$ represents an abstract relationship ``$\lhd$", if its vertexes
$v_i$ represent elements of a set, and each link $e_{ij}$ represents a $\lhd$-related pair, $v_i \lhd v_j$.


\vspace{10 pt}
\textbf{Simplicial complexes} 
\vspace{7 pt}

\vspace{3 pt}
$\bullet$ \textit{Geometric simplexes} are points ($0$-simplexes, $\kappa^{(0)}$), line segments ($1$-simplexes,
$\kappa^{(1)}$), triangles ($2$-simplexes, $\kappa^{(2)}$), tetrahedra ($3D$-simplexes, $\kappa^{(3)}$), as well
as their $d>3$-dimensional generalizations (Fig.~\ref{fig:simplx}A). Note that the set of vertexes opposite to
a given vertex in a $d$-simplex $\kappa^{(d)}$ spans a $(d-1)$-simplex---a \textit{face} of $\kappa^{(d)}$. The
boundary of a $d$-simplex then consists of $(d+1)$ faces $\kappa^{(d-1)}_1,\kappa^{(d-1)}_2,\ldots,
\kappa^{(d-1)}_{d+1}$ (Fig.~\ref{fig:simplx}B).

\vspace{3 pt}
$\bullet$ \textit{Geometric simplicial complexes} are combinations of geometric simplexes that match each 
other vertex-to-vertex, so that a non-empty intersection of any two simplexes in $K$ yields another $K$-simplex:
if $\kappa_1,\kappa_2\in K$, then $\kappa_1\cap\kappa_2=\kappa_3\in K$. 

\vspace{3 pt}
$\bullet$ The collection of all simplexes of dimensionality $d$ and less forms the \textit{$d$-skeleton} of $K$,
$sk_d(K)$.

\vspace{3 pt}
$\bullet$ Topological analyses of simplicial complexes do not address simplexes' shapes and are based entirely
on the combinatorics of the vertexes shared by the simplexes. This motivates using \textit{abstract simplexes}
and \textit{abstract simplicial complexes} that capture the combinatorial structure of $\kappa^{(d)}$s without
making references to their geometry. Specifically, an abstract $0$-simplex is a vertex $\sigma^{(0)}_i\equiv 
v_i$, an abstract $1$-simplex is a pair of vertexes, $\sigma^{(1)}_{ij}=[v_i,v_j]$; an abstract $2$-simplex is
a triple of vertexes, $\sigma^{(2)}_{ijk}=[v_i,v_j,v_k]$, and so forth (Fig.~\ref{fig:simplx}C). Thus, abstract
complexes may be viewed as multidimensional generalizations of graphs or as abstractions derived from the 
geometric simplicial complexes. 

A $d$-element subset of an abstract $d$-simplex $\sigma^{(d)}$ forms its $(d-1)$-face. The``face-matching" of
the abstract simplexes in $\Sigma$ means simply that a nonempty overlap of two simplexes $\sigma_1,\sigma_2\in
\Sigma$ is a simplex of the same complex, $\sigma_1\cap\sigma_2=\sigma_3\in\Sigma$. The latter property is 
commonly used to define abstract simplicial complexes for arbitrary sets, using families of their subsets that
are closed under the ``$\cap$" operation \cite{Alexandrov}.

\begin{figure}[h]
	\includegraphics[scale=0.8]{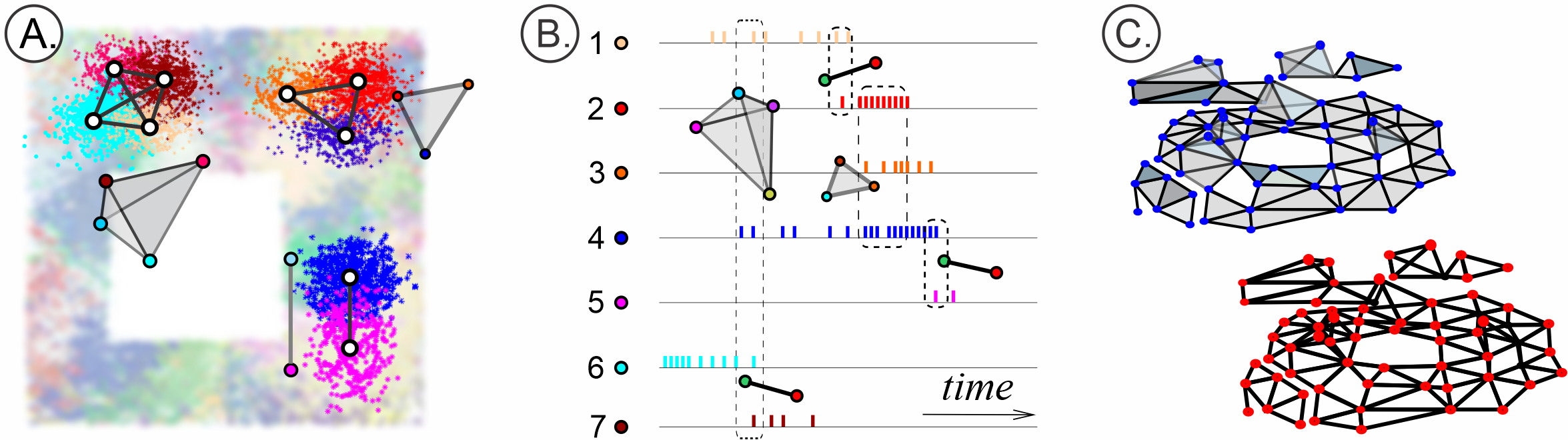}
	\caption{\footnotesize \textbf{Cliques and simplexes}.
		(\textbf{A}). Pairwise interlinked place fields produce cliques of the coactivity graph $\mathcal{G}$.
		Shown is a vertex $\varsigma^{(0)}$ ($0$-clique), a link $\varsigma^{(1)}$ ($1$-clique), a three-vertex 
		$\varsigma^{(2)}$ and a four-vertex $\varsigma^{(3)}$ clique. 
		(\textbf{B}). Geometric simplexes: a $0D$ dot ($\kappa^{(0)}$), a $1D$ link ($\kappa^{(1)}$), a $2D$
		triangle ($\kappa^{(2)}$) and a $3D$ tetrahedron ($\kappa^{(3)}$).
		(\textbf{C}). The corresponding complexes: a simplicial coactivity complex $\mathcal{T}_{\sigma}$ 
		whose simplexes (\ref{sigma}) are detected as singular coactivity events (left) may topologically 
		differ from the clique coactivity complexes $\mathcal{T}_{\varsigma}$, assembled from the cliques
		of a coactivity graph $\mathcal{G}$ (right) over a spike integration period $\varpi$. A simplicial
		complex $K$ is a combination of matching simplexes. The set of vertexes and black lines highlight
		the $1D$-skeleton of $sk_1(K)$. 
	} 
	\label{fig:complx}
\end{figure}

\begin{itemize}[nosep]
\item \textit{Example 1}: The set of overlapping regions (\ref{overlap}) define abstract simplexes (\ref{nsimplex})
of the nerve complex (\ref{nerve}) (Fig.~\ref{fig:complx}A).
\item \textit{Example 2}: The combinations of coactive cells define coactivity simplexes (\ref{sigma}),
which together form a coactivity complex (Fig.~\ref{fig:complx}B).
\item \textit{Example 3}. Vertexes of geometric simplexes that form a geometric simplicial complex $K$ define
abstract simplexes that form the corresponding abstract simplicial complex $\Sigma$ (Fig.~\ref{fig:complx}C).
\end{itemize}

\vspace{3 pt}
$\bullet$ The set of $d$-dimensional simplexes of a complex $\Sigma$ forms its (abstract) $d$-skeleton, $sk_d(\Sigma)$.

\vspace{3 pt}
$\bullet$ 
\textit{A clique complex} of an undirected graph $G$ is an abstract simplicial complex formed by the cliques
(fully interconnected subgraphs) of $G$ \cite{Jonsson}, Fig.~\ref{fig:simplx}A. Combinatorial properties of
cliques are the same as simplexes': a subset of a clique's vertexes form a clique, overlap of two $G$-cliques
is also a clique, $\varsigma_1\cap\varsigma_2=\varsigma_3\in G$ (Fig.~\ref{fig:simplx}). Thus, any graph $G$
defines a unique clique complex $\tilde{\Sigma}(G)$. Note, that the $1$-skeleton of a clique complex yields
its underlying graph, $\sk_1(\tilde{\Sigma}(G))=G$, but if $\Sigma$ is not a clique complex, then the clique
complex built over its $1$-skeleton does not reproduce $\Sigma$. 

\vspace{3 pt}
$\bullet$ \textit{Coactivity complexes} used in this study are 
of two kinds. The first kind is formed by the abstract complexes $\mathcal{T}_{\sigma}$ built from simultaneously 
coactive cell groups (\ref{sigma}). The second kind is formed as the clique complexes of the coactivity graphs
$\mathcal{G}$ \cite{CAs,Hoffman}. The graph (co)activity thresholds $m$ and $\mu$ are used to control the size
of the complex $\mathcal{T}_{m,\mu}=\mathcal{T}(\mathcal{G}_{m,\mu})$: selecting $m\geq 2$, $\mu =1$ for small
maps (i.e., counting cells that produce at least two spikes per time bin $w$) and $m\geq 2$, $\mu\geq 5$ for 
larger maps allows computing the full simplicial complex with dimensionality $\dim(\mathcal{T}_{m,\mu})\leq 10$,
for which we can numerically apply the \textsf{Javaplex} software \cite{javaplex}.

\vspace{10 pt}
\textbf{Topological invariants}.
\vspace{7 pt}

\vspace{3 pt}
$\bullet$ \textit{Homological groups} are designed to ``count pieces" in a space $X$ with suitable coefficients.
The key property of these groups is that they remain unchanged---\textit{invariant}---as $X$ is continuously
deformed (see \cite{Hatcher,Alexandrov} for a gentle introduction to the subject). If the coefficients form an
algebraic field $F$, then the homological groups, commonly referred to as the ``homologies" of $X$ are simply
vector spaces $H_{0}(X,F),H_{1}(X,F),\ldots$, associated with $X$ (one per dimension). Homologies can be easily
computed for spaces whose ``pieces" are explicitly defined, e.g., for the simplicial complexes, thus providing
a way of identifying their topological structures. In practice, it is easier to use just the dimensionalities 
of $H_{\ast}$s---the \textit{Betti numbers} $b_k= \dim(H_k(X,F))$, to count numbers of connectivity components,
cavities, tunnels and other topological features of $X$ in different dimensions \cite{Alexandrov,Hatcher}. For
example, if $X$ is the boundary of a hollow triangle (or another noncontractible $1D$ loop), then $\beta_1(X)=1$.
If $X$ is $1$-dimensional complex, i.e., a graph, then $\beta_1(X)$ equals to the number  of cycles in $X$, 
counted up to topological equivalence. If the triangle is ``filled", then it can be continuously contracted 
into a $0D$ point; since the latter has no topological structure in dimensions $d>0$, the corresponding Betti
numbers also vanish. By the same argument a ``filled" tetrahedron has $\beta_{k>0}=0$, but if the tetrahedron
is hollow, then its boundary, being a $2D$ noncontractible loop (topologically---a $2D$ sphere) produces 
$\beta_2=1$, $\beta_{k>2}=0$. Similarly, for any $d$-simplex $\beta_{k>0}(\sigma^{(d)}) = 0$, whereas for its
hollow boundary, $\partial\sigma^{(d)}$, the Betti numbers are $\beta_{d-1}(\partial\sigma^{(d)})=1$, 
$\beta_{k\neq 0,d-1}(\partial\sigma^{(d)})=0$ (Fig.~\ref{fig:simplx}). Same results apply to the ``abstract"
counterparts of all these complexes. 
Note also that continuous deformations of a $0D$ point $x$ (a $0D$ topological loop) amount to ``sliding" $x$
inside of a space $X$ that contains $x$; thus $\beta_0(X)$ simply counts such ``sliding domains", i.e., the 
number of connected components in $X$. As a result, all simplexes and simplicial complexes that consist of one
piece have $\beta_0(X)=1$.

\vspace{3 pt}
$\bullet$ \textit{Persistent homology} theory allows tracing the topological structure in a filtered family
of simplicial complexes, e.g., describing the topological dynamics of the inflating family (\ref{complex}),
\cite{EdelZom,Wasserman,Zomorodian,ZomorodianBook}. The Betti numbers plotted as function of the filtration
parameter (in our case it is time, $t$) form the \textit{barcode}, $\mathfrak{b}(\mathcal{T},t)=
(b_0(\mathcal{T},t),b_1(\mathcal{T},t),\ldots)$, which provides the exact mathematical meaning to the term 
``topological shape" used throughout the text. Each bar in $\mathfrak{b}(\mathcal{T},t)$ can be viewed as 
the corresponding topological loop's timeline \cite{PLoS,Arai,Basso,Hoffman,CAs,Eff}. 

\vspace{3 pt}
$\bullet$ 
\textit{Zigzag Homology} theory allows tracking the Betti numbers of the ``flickering" complexes---the ones
whose simplexes can not only appear, but also disappear (see \cite{Carlsson1,Carlsson2} and Supplement in 
\cite{PLoZ}). In particular, Zigzag homology techniques allow capturing the times when individual loops appear
in the flickering complex, how long they persist, when they disappear, reappear again, etc.

\vspace{10 pt}
\textbf{Representability}
\vspace{7 pt}

A generic algorithm for checking whether a given complex can be built as a nerve of a $D$-dimensional cover
is known only for $D=1$ (see below). However, there exist criteria that allow ruling out certain non--representable
cases.

\vspace{3 pt}
$\bullet$ The \textit{Leray criterion} posits that if a complex $\Sigma$ is a nerve of a $D$-dimensional cover 
with contractible overlaps (\ref{overlap}), then its rational homologies in dimensions higher or equal than $D$ should 
vanish, $H_{i\geq D}(\Sigma,\mathbb{Q})=0$ \cite{Leray}. Moreover, homologies of all the subcomplexes $\Sigma'
\subseteq \Sigma$, induced by selecting vertex subsets of $\Sigma$ should also vanish, $H_{i\geq D}(\Sigma',
\mathbb{Q})=0$. 
These properties can be verified by computing the Betti numbers and verifying that $b_{i\geq D}(\Sigma,\mathbb{Q})=0$. 
In practice, it is more convenient to carry out the computations over a finite field, such as $\mathbb{Z}_2$. 
Although the $b_k(\Sigma,\mathbb{Q})$ numbers may in general differ from the $b_k(\Sigma,\mathbb{Z}_2)$ numbers,
the latter also have to obey the Leray condition and produce the same Leray dimensionality. As an example, the
Leray condition poses that the boundary of the triangle is not $1$-representable ($\beta_1(\partial\sigma^{(2)},
\mathbb{Z}_2)>0$), but the triangle itself may be ($\beta_1(\sigma^{(2)},\mathbb{Z}_2)=\beta_2(\sigma^{(2)},\mathbb{Z}_2)=0$); the boundary of a 
tetrahedron is not $2$-representable ($\beta_2(\partial\sigma^{(3)},\mathbb{Z}_2)>0$), but the tetrahedron may be.

\begin{wrapfigure}{c}{0.5\textwidth}
	\includegraphics[scale=0.66]{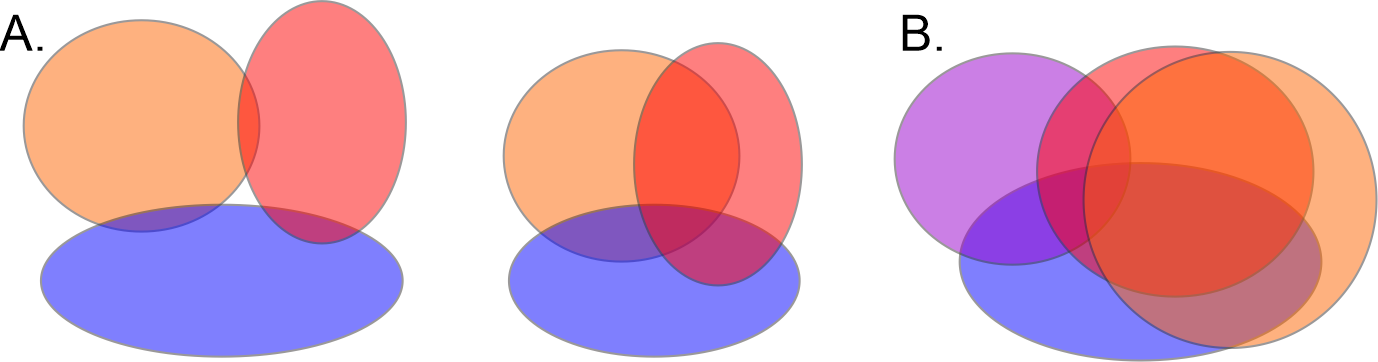}
	\caption{{\footnotesize
			\textbf{Helly's theorem}. (\textbf{A}). Three regions may exhibit both pairwise (left) or triple
			overlap (right). (\textbf{B}). The Helly number of a family of convex regions in $\mathbb{R}^d$ does not exceed
			$d+1$. Thus, for convex planar regions having all triple overlaps implies having all the higher order (i.e., for
			this particular picture quadruple) overlaps. Hence, the intersection patterns of convex planar subspaces are 
			completely determined by the intersection patterns of triples.  
	}}
	\label{fig:helly}
\end{wrapfigure}

\vspace{3 pt}
$\bullet$ \textit{Amenta's theorem} connects the Leray dimensionality of a simplicial complex to its 
\textit{Helly number}, defined as follows. Let $\Upsilon=\{\upsilon_1,\upsilon_2,\ldots,\upsilon_n\}$ be a
finite family of regions (Fig.~\ref{fig:helly}). The Helly number $h=h(\Upsilon)$ of the family is defined
to be the maximal number of non-overlapping regions, such that every $h-1$ among them overlap. 
For the corresponding nerve complex $\mathcal{N}(\Upsilon)$, $h=h(\mathcal{N})=h(\Upsilon)$ is the number 
of vertices of the largest simplicial hole in $\mathcal{N}$ (i.e., the dimension of the hole plus $2$,
\cite{TancerSur}). This observation can be used to attribute a Helly number to any simplicial complex 
$\Sigma$, $h(\Sigma)$. 
From the perspective of representability analyses, a key property of the Helly numbers is that they do not
exceed $d+1$ for a $d$-Leray complex \cite{TancerSur}. In particular, if the regions $\upsilon_i\in\Upsilon$ 
consist of up to $k$ compact, convex domains in $R^d$, and any intersection $\upsilon_{i_1} \cap \dots \cap 
\upsilon_{i_t}$ also satisfies this property, then $h(\Upsilon)\leq k(d+1)$ \cite{Amenta,TancerSur,Danzer}.

\vspace{3 pt}
$\bullet$ \textit{Eckhoff's conjecture}. The $f$-vector $f=(f_1,f_2,\ldots,f_n)$ of a simplicial complex $\Sigma$ 
is the list of numbers of its $k$-dimensional simplexes, $f_k=\#\{\sigma_i\in\Sigma|\dim(\sigma)=k\}$ (``$f$" 
is a traditional notation that should not be confused with the firing rates). The $h$-vector of $\Sigma$ is
defined as
\begin{equation}
h_k= \begin{cases}
f_k, \,\,\, & \mbox{for} \,\,\,k=0,1,\ldots,D-1, \\ 
\sum_{j\geq 0} (-1)^j  {{k+j-D}\choose{j}} f_{k+1}, \,\,\, & \mbox{for} \,\,\, k=D,D+1,\ldots\, ,
\end{cases}
\nonumber
\end{equation} 
where parentheses denote the binomial coefficients. Given the combinatorial decomposition of $l$,
\begin{equation}
l={{l_k}\choose{k}}+{{l_{k-1}}\choose{k-1}}+\ldots+{{l_j}\choose{j}},
\end{equation}
where $l_k\geq l_{k-1}\geq\ldots\geq l_j\geq j\geq 1$ \cite{Beckenbach}, define the set of numbers
$$l^{(k)}={{l_k}\choose{k-1}}+{{l_{k-1}}\choose{k-2}}+\ldots+{{l_j}\choose{j-1}},$$ with $0^{(k)}=0$. 
Eckhoff's conjecture \cite{Eckhoff}, proven in \cite{Kalai1} holds that the $h$-numbers of
a $d$-representable complex must satisfy the following inequalities:
\begin{equation}
\begin{cases}
h_k\geq 0 & \mbox{for} \,\,\, k=0,1,\ldots; \\
h^{(k+1)}_k \leq h_{k-1}, \,\,\, & \mbox{for} \,\,\,k=1,2,\ldots,D-1; \\ 
h^{(d)}_k\leq h_{k-1}-h_k, \,\,\, & \mbox{for} \,\,\, k=D,D+1,\ldots\,.
\end{cases}
\nonumber
\end{equation} 
which can be verified not only for ``static" complexes, but also for the ``inflating" (\ref{complex}) and ``flickering"  
complexes, at each step of their evolution.

\vspace{10 pt}
$\bullet$ \textit{Qualitative spatial consistency}. It can be shown that if the \textsf{RCC5} relationships
among all triples of regions are consistent, then the entire schema $\mathcal{R}_5$ is consistent \cite{Renz,
Cohn94g,CohnRenz,ChenCohn,Bennet1,Long}. The full set of consistent triples is given in the following table.

\begin{table}[ht]
\centering
\begin{tabular}{|c||c|c|c|c|c|}
\hline
$\circ$ & $\mathsf{DR}(y,z)$ & $\mathsf{PO}(y,z)$ & $\mathsf{PP}(y,z)$ & $\mathsf{PPi}(y,z)$ & $\mathsf{EQ}(y,z)$ \\ \hline\hline
$\mathsf{DR}(x,y)$ & $\mathsf{any}$ & $\mathsf{DR, PO,PP}$ & $\mathsf{DR, PO,PP}$ & $\mathsf{DR}$ & $\mathsf{DR}$ \\ \hline
$\mathsf{PO}(x,y)$ & $\mathsf{DR, PO,PPi}$ & $\mathsf{any}$ & $\mathsf{PO, PP}$ & $\mathsf{DR, PO,PPi}$ & $\mathsf{PO}$ \\ \hline
$\mathsf{PP}(x,y)$ & $\mathsf{DR}$ & $\mathsf{DR, PO,PP}$ & $\mathsf{PP}$ & $\mathsf{any}$ & $\mathsf{PP}$ \\ \hline
$\mathsf{PPi}(x,y)$ & $\mathsf{DR, PO,PPi}$ & $\mathsf{PO, PPi}$ & $\mathsf{PO, EQ,PP,PPi}$ & $\mathsf{PPi}$ & $\mathsf{PPi}$ \\ \hline
$\mathsf{EQ}(x,y)$ & $\mathsf{DR}$ & $\mathsf{PO}$ & $\mathsf{PP}$ & $\mathsf{PPi}$ & $\mathsf{EQ}$ \\ \hline
\end{tabular}
\caption{{\footnotesize
\textbf{$\mathsf{RCC5}$ compositions}. Given three regions, $x$, $y$ and $z$, and two relationships
$\mathsf{R}_1(x,y)$ and $\mathsf{R}_2(y,z)$, the relationship $\mathsf{R}_3(x,z)$ is not arbitrary.
A map is consistent, if every triple of relationships is $\mathsf{RCC5}$--consistent.}}
\label{tab:comp}
\end{table}

\vspace{3 pt}
$\bullet$ \textit{Recognizing $1$-representability} algorithm follows the exposition in \cite{Golumbic,Fishburn}.

Let $\mathcal{I}=\{I_1=[a_1,b_1],\ldots,I_n=[a_n,b_n]\}$ be set of intervals of a Euclidean line $R^1$.

\underline{Definition 1}. \textit{$G(\mathcal{I})$ is an interval graph, if each vertex $v_i\in G(\mathcal{I})$
corresponds to an interval $I_i\in\mathcal{I}$ and a pair of vertexes $(v_i,v_j)$ is connected by an edge iff
$I_i$ and $I_j$ intersect.}

An interval graph is hence $1$-dimensional skeleton of the nerve of $\mathcal{I}$ (Fig.~\ref{order}A). It can
also be verified that the complement of an interval graph is a directed comparability graph $\tilde{G}_{\lhd}
(\mathcal{I})$, in which the relationship $v_i\lhd v_j$ is defined by the order of the overlapping intervals,
\begin{equation}
v_i\lhd v_j \implies b_i<a_j.
\label{order}
\end{equation}

\underline{Definition 2}. \textit {A directed graph satisfies $\Lambda$-property if there are no three vertices
	$v_i,v_j,v_k$ such that $v_i,v_k$ are not adjacent, while $v_i$ is adjacent to $v_j$ and $v_j$ is adjacent
	to $v_k$ with the corresponding orientations being $e_{ij}$ and $e_{jk}$ respectively}.

\underline{Definition 3}. \textit{An interval graph satisfies \textit{$\times$-property} if no four vertexes 
$v_i,v_j, v_k, v_l$ produce disjoint pairs of intervals}. 

In other words, a situation when  $v_i\lhd v_j$ and $v_k\lhd v_l$ (i.e., the pair of intervals $(I_i,I_j)$ 
overlaps and the pair $(I_k,I_l)$ also overlaps), while the remaining pairs remain incompatible, e.g., $v_j
\ntriangleleft v_k$, $v_j \ntriangleleft v_l$, (i.e., $I_j$ does not overlap either $I_k$ or $I_l$), etc., 
does not appear.

\underline{Theorem}. {\it A graph is an interval graph iff its complement is a comparability graph with an 
order defined by (\ref{order}), satisfying the $\times$-property.}

\vspace{3 mm}
This theorem and the definitions motivate the following algorithm for identifying $1D$ representability of a
complex $\Sigma$ (Fig.~\ref{fig:alg}B):

\begin{figure}[h]
	\includegraphics[scale=0.84]{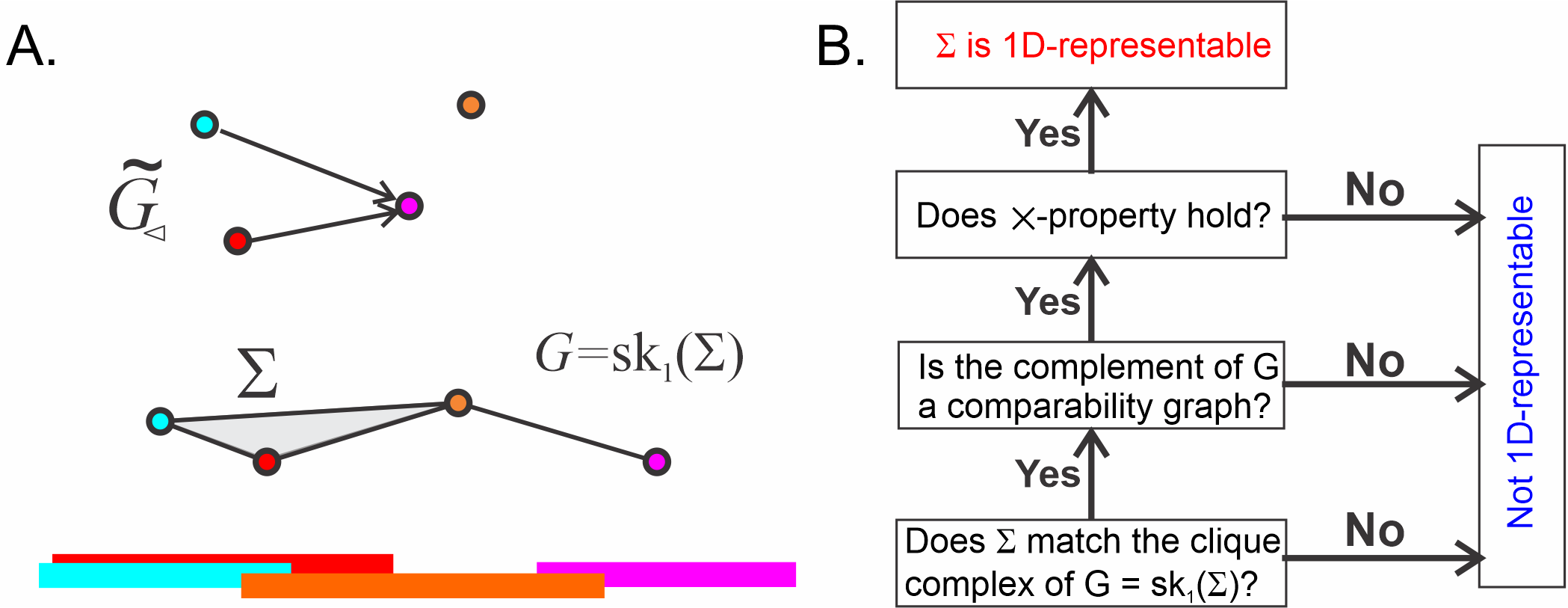}
	\caption{{\footnotesize
			\textbf{An algorithm for recognizing $1$-representable complexes}. 
			\textbf{A}. Four intervals covering a linear segment (bottom) can be represented by a simplicial
			complex---the nerve of the cover (middle panel). The vertexes of the corresponding interval graph
			$G(\mathcal{I})$---the $1D$ skeleton of $\Sigma$---(color-coded) are connected if their respective
			intervals overlap, $I_i \cap I_j\implies v_i\lhd  v_j$. The corresponding comparability graph,
			$\tilde{G}_{\lhd}(\mathcal{I})$ is shown above, with the order indicated by arrows: $v_i \lhd v_j$
			iff there is an arrow leading from $v_i$ to $v_j$.
			\textbf{B}. Given a simplicial complex
			$\Sigma$, first check whether it is the clique complex of its $1D$-skeleton $G:=sk_1(\Sigma)$.
			If it is not, then $\Sigma$ is not $1$-representable; if it is, then check whether the complement
			graph of $G$ is a comparability graph. If it is not, then $\Sigma$ is not $1-$representable. If it
			is, then check the $\times$-property: if it holds, then $\Sigma$ is $1$-representable, otherwise it is 
			not. 
	}}
	\label{fig:alg}
\end{figure} 

\vspace{3 pt}
\textbf{1}. Test whether $\Sigma$ is a clique complex, i.e., verify whether all $(k+1)-$tuples of vertexes
$v_{\sigma}=[v_{i_0},v_{i_1},\ldots,v_{i_k}]$ form a simplex in $\Sigma$ if and only if each pair of vertexes
$[v_{i_p},v_{i_q}]\in v_{\sigma}$ is an edge in its $1$-skeleton $G=sk_1(\Sigma)$. If at least one 
$v_{\sigma}$ fails this test, then $\Sigma$ is not a clique complex and hence not representable.

\textbf{2}. Build the complement $\tilde{G}$ of $\sk_1(\Sigma)$ and verify its comparability as follows:

\textit{i}. Choose an edge between $v_i$ and $v_j$ and define an orientation on it (e.g., $e_{ij}\neq e_{ji}$).
If $e_{ij}$ was selected, then search for all vertexes $v_{j'}$ that are connected to $v_{j}$ but not to $v_{i}$
(Fig.~\ref{fig:alg}A). If the edge between $j$ and $j'$ is not yet oriented, select $e_{j'j}$. If it was already 
$(j'j)$-oriented, continue on; the opposite, $(jj')$-orientation implies that $\Sigma$ is not  representable. 

If the orientation for new edges cannot be selected, pause the algorithm and dispose of all the edges that
have already been oriented. Then pick another unoriented edge and restart the $\Lambda$-rule: keep applying
it until the process comes to a halt and the next  set of edges needs to be removed. Do this until all the 
edges are serviced and hence oriented.

\textit{ii}. Verify that no $3$-tuple of vertexes $(v_i,v_j,v_k)$ forms an oriented $3$-cycle. If such a cycle 
exists, $\Sigma$ is non-representable in $1D$. 

\textit{iii}. Verify that no triple of vertexes $(v_i,v_j,v_k)$ is ``disconnected," i.e., given $e_{ij}$ and
$e_{jk}$, there must exist an edge between $i$ and $k$. If any triple violates this condition, $\Sigma$ is not
representable. Otherwise $\tilde{G}=\sk_1(\Sigma)$ is a comparability graph with the order: $v_i\lhd  v_j$ for
each $e_{ij}$.

\textbf{3}. For every vertex $v_i$, compute the set of lesser points, $D(v_i)=\{v_j \in V: v_j\lhd v_i\}$. Then,
for all pairs of vertexes $(v_i,v_j)$ check whether $D(v_i)$ is a subset of $D(v_j)$ or vice-versa. If at least
one of these conditions is not satisfied, $\Sigma$ is not representable. 

If this sequence of conditions is satisfied, $\Sigma$ is $1D$-representable.



\clearpage
\newpage

\section{References}
\label{sec:refs}

\end{document}